\documentclass[
    aps,
    pra,
    reprint,
    superscriptaddress,
    floatfix,
    nofootinbib
]{revtex4-2}

\usepackage{amsmath}
\usepackage{amssymb}
\usepackage{graphicx}
\usepackage{mathrsfs}
\usepackage{microtype}
\graphicspath{{figures/}}
\usepackage[colorlinks=true, linkcolor=blue, citecolor=blue, urlcolor=blue]{hyperref}
\begin{document}

\title{Recalibration-Aware Optimization of ZEFOZ Operating Points in
\texorpdfstring{$^{167}\mathrm{Er}^{3+}:\mathrm{Y}_2\mathrm{SiO}_5$}{167Er3+:Y2SiO5}
for Quantum Memories}

\author{Shashank Srivastava}
\affiliation{Department of Physics, Netaji Subhas University of Technology,
New Delhi 110078, India}

\begin{abstract}
Erbium-doped solids like $^{167}\mathrm{Er}^{3+}:\mathrm{Y}_2\mathrm{SiO}_5$ have properties that make them suitable for quantum-network memories, however, their unquenched effective electronic spin causes rapid magnetic decoherence. Operating at Zero First-Order Zeeman (ZEFOZ) points allows us to mitigate this, where we evaluate the ZEFOZ landscape for Site-2 $^{167}\mathrm{Er}^{3+}$ by incorporating a full-Hessian spin-Hamiltonian treatment with the anisotropic magnetic-field covariance derived from the native $^{89}\mathrm{Y}$ crystal lattice directly. Operating-point screening criteria were adopted to devise a Pareto-optimal frontier of four transition pairs that was targeted to explicitly balance modeled Y-only magnetic pure-dephasing lifetimes ($T_{\phi,Y}^{(2)}$) against microwave control authority. We noted that the $\epsilon_{14}\leftrightarrow\epsilon_{15}$ transition exhibits the longest modeled Y-bath-induced quadratic pure-dephasing time, $T_{\phi,Y}^{(2)}>2.2\times10^5~\mathrm{s}$, and for an optimized hardware addressability the $\epsilon_{12}\leftrightarrow\epsilon_{13}$ transition stood out, achieving a Rabi frequency coefficient of $1.248 \times 10^5~\mathrm{Hz/mT}$. Furthermore, upon executing an assumption-based perturbed-parameter Hamiltonian-uncertainty analysis, the results demonstrate that while uncompensated deviations shift the ZEFOZ condition away from its nominal magnetic-field coordinate, the scope for successfully recovering the shifted operating points via numerical vector-field recalibration remains entirely viable. The Pareto-optimal coordinates were mathematically recovered for all the sampled perturbations that were inputted in the Hamiltonian, and the $84.1^{\mathrm{st}}$-percentile field relocations ranged from $10.95$ to $27.97~\mathrm{mT}$. The intended operation near $10~\mathrm{mK}$ must be taken into account for the suppression of Er-induced spectral diffusion through electronic-spin polarization, while the extended Y-bath pure-dephasing timescales also remain conditional on the adopted temporal-correlation model. 
\end{abstract}

\maketitle
\section{Introduction}
\label{sec:introduction}

\subsection{Motivation: Erbium-Based Spin-Wave Quantum Memories}
\label{subsec:motivation}

Erbium doped solids possess properties which make them suitable quantum-network applications \cite{kimble2008,afzelius2010}. Their optical transitions lie in the telecommunication C-band \cite{jiang2023,rancic2018} which can be harnessed for quantum-network memories, especially \textbf{$^{167}\mathrm{Er}^{3+}:\mathrm{Y}_2\mathrm{SiO}_5$}. However, it being a Kramers rare-earth ion, the magnetic environment poses major hurdles for long-lived spin-wave storage. $\mathrm{Er}^{3+}$ has an effective electronic spin $S=1/2$ in its ground-state Kramers doublet \cite{wang2023,g23x-14ks} which produces a large electronic magnetic moment. Due to which the available storage time is severely limited as the hyperfine transition becomes sensitive to fluctuating magnetic fields generated by the surrounding nuclear and electronic spin baths \cite{bottger2006,rakonjac2020,ulanowski2025}.

\subsection{ZEFOZ Protection and Its Limits}
\label{subsec:zefoz-limits}

Operating at a Zero First-Order Zeeman (ZEFOZ) point is an effective method to suppress magnetic sensitivity which removes the first-order response to magnetic field fluctuations by eliminating the transition-frequency gradient ($\mathbf{S}_1=\mathbf{0}$) \cite{fraval2004,mcauslan2012}. The transition remains sensitive to second order-order perturbations, by which the resulting quadratic magnetic pure dephasing depends not only on the frequency Hessian but also on the spatial distribution of the bath-induced, temporally-fluctuating magnetic fields\cite{mcauslan2012,kubo1962}.

\subsection{Relation to Prior ZEFOZ and Bath Studies}
\label{subsec:prior-work}

Matsuura et al.~\cite{g23x-14ks} have already laid the foundation work for this methodology by identifying ZEFOZ transitions in  $\mathrm{Er}^{3+}:\mathrm{Y}_2\mathrm{SiO}_5$ using a tensorial formulation. Here, we further treat the magnetic bath by calculating the anisotropic $^{89}\mathrm{Y}$ field covariance directly from the crystal and contracting it with the complete transition-frequency Hessian. This approach enables us to evaluate the quadratic frequency-noise variance while taking the directional contributions of both tensors into account. A temporal-correlation model is then introduced to estimate the resulting magnetic pure-dephasing timescale. 

\subsection{Scope and Contributions}
\label{subsec:scope}

We probe into conditional Y-bath to measure the induced quadratic magentic pure-dephasing time , $T_{\phi,Y}^{(2)}$, for Site-2 $^{167}\mathrm{Er}^{3+}:\mathrm{Y}_2\mathrm{SiO}_5$. The calculation clubs together a full-Hessian second cumulant treatment with lattice-derived magnetic-field covariance, in addition with an assumed exponential correlation function. We begin with the numerically identified ZEFOZ roots and subject them to physical screening criteria to then isolate four Pareto-optimal transition pairs that efficiently manage both magnetic protection with microwave addressability. In a later approach, we examine how uncertainties in the spin-Hamiltonian parameters shift these operating points, to compare the extent of recovery achievable via open-loop operation and numerical magnetic-field recalibration scenarios.

\section{Model and Computational Methods}
\label{sec:methods}

\subsection{Site-2 Effective Spin Hamiltonian}
\label{subsec:methods_hamiltonian}

$^{167}\mathrm{Er}$ has a nuclear spin of $I=7/2$ and this isotope has a natural abundance of approximately $22.9\%$ \cite{nistEr}. The $4f^{11}$ ground-state manifold of $^{167}\mathrm{Er}^{3+}$ substituted at Site 2 of $\mathrm{Y}_2\mathrm{SiO}_5$ is modeled. When the temperature is in the range of 10 mK and 4 K, there is not much thermal energy to create excited crystal-field doublets and we can safely treat the system as a pure effective electronic spin $S=1/2$ \cite{wang2023,g23x-14ks}. As the electronic ground-doublet, represented by an effective spin $S=1/2$, combines with $I=7/2$ nuclear spin, it gives a composite Hilbert space $\mathscr{H} =\mathbb{C}^{2} \otimes \mathbb{C}^{8}$, with dimension: 
\begin{equation}
\dim(\mathscr{H}) = (2S + 1)(2I + 1) = 16.
\label{eq:hilbert_dim}
\end{equation}
The effective Hamiltonian is expressed as $\mathcal{H} = H/h$, such that its eigenvalues are level frequencies in MHz, and we easily obtain the transition frequencies by the difference of these eigenvalues. The Hamilton incorporates four interactions:
\begin{equation}
\mathcal{H} = \mathcal{H}_{\mathrm{hf}} + \mathcal{H}_{\mathrm{Q}} + \frac{\mu_B}{h} \mathbf{B} \cdot \mathbf{g}_0 \cdot \mathbf{S} - \frac{g_n \mu_N}{h} \mathbf{B} \cdot \mathbf{I}.
\label{eq:hamiltonian_freq}
\end{equation}
Here $\mu_B/h = 13.996245\text{ MHz/mT}$ is the Bohr Magneton coefficient and $g_n \mu_N/h \simeq -1.23 \times 10^{-3}\text{ MHz/mT}$ is the adopted Nuclear Zeeman coefficient~\cite{wang2023,mcauslan2012}.

The dimensionless spin operators  $s_\alpha$ and $i_\alpha$ act on the electronic and nuclear subspaces, respectively.
Their representations on the composite space are:

\begin{equation} 
\widetilde{S}_\alpha = s_\alpha \otimes \mathbf{I}_8, \qquad \widetilde{I}_\alpha = \mathbf{I}_2 \otimes i_\alpha, \label{eq:composite_ops} 
\end{equation}
where $\alpha \in \{D_1, D_2, b\}$ aligns with the mutually orthogonal optical dielectric frame. The hyperfine ($\mathbf{A}_0$) and nuclear quadrupole ($\mathbf{Q}_0$) interactions take the form \cite{wang2023,sun2008}:
\begin{align}
\mathcal{H}_{\mathrm{hf}} &= \sum_{\alpha,\beta} (A_0)_{\alpha\beta} \, s_\alpha \otimes i_\beta, \label{eq:hf_term} \\
\mathcal{H}_{\mathrm{Q}} &= \sum_{\alpha,\beta} (Q_0)_{\alpha\beta} \, \mathbf{I}_2 \otimes i_\alpha i_\beta. \label{eq:q_term}
\end{align}

Since $\mathrm{Y}_2\mathrm{SiO}_5$ has a two-fold $C2$ rotation axis, it belongs to the monoclinic space group ($C2/c$) \cite{sun2008,g23x-14ks}. 
Erbium gets substituted in two distinct crystallographic sites in the unit cell. They are distinct by the local point group symmetry , which is $C_1$ at these substituted sites. The symmetry-related magnetic subsite obtained by the crystallographic $b$ axis twofold rotation is selected by the rotation matrix $R_{C_2} = \operatorname{diag}(-1, -1, 1)$. The axis is represented in $(D_1, D_2, b)$ frame. The tensors used for this calculation are:

\begin{equation}
T_0 = R_{C_2} T_{\mathrm{site2}} R_{C_2}^{\mathsf{T}}, \qquad T \in \{A, Q, g\},
\label{eq:c2_transform}
\end{equation}

where the empirical Site-2 tensors are adopted from Ref.~\cite{wang2023} and reproduced in Appendix~\ref{app:hamiltonian}.

Upon diagonalizing the the Hermitian operator $\mathcal{H}(\mathbf{B})$, we obtain 16 real eigenvalues. They are sorted and labeled as:
\begin{equation}
\epsilon_0(\mathbf{B}) \le \epsilon_1(\mathbf{B}) \le \dots \le \epsilon_{15}(\mathbf{B}).
\label{eq:energy_order}
\end{equation}
A transition frequency between an upper level $j$ and a lower level $i$ ($j > i$) is defined as:
\begin{equation}
f_{ij}(\mathbf{B}) = \epsilon_j(\mathbf{B}) - \epsilon_i(\mathbf{B}).
\label{eq:transition_freq}
\end{equation}
A zero-based eigenstate indexing is maintained throughout subsequent derivations.

\subsection{Automatic Differentiation of Transition Frequencies}
\label{subsec:methods_autodiff}

To preserve information, the energy levels must not fluctuate. A local Taylor expansion about an operating point $B_0$ is used to measure the arbitrary hyperfine transition frequency $f_{ij}(\mathbf{B})$ response to small applied magnetic field fluctuations  $\delta\mathbf{B} = (\delta B_{D_1}, \delta B_{D_2}, \delta B_b)^{\mathsf{T}}$. Subject to local smoothness, 

\begin{align}
f_{ij}(\mathbf{B}_0+\delta\mathbf{B})
={}& f_{ij}(\mathbf{B}_0)
+\mathbf{S}_1(\mathbf{B}_0)^{\mathsf T}\delta\mathbf{B}
\nonumber\\
&+\frac12\delta\mathbf{B}^{\mathsf T}
\mathbf{S}_2(\mathbf{B}_0)\delta\mathbf{B}
+\mathcal{O}(\|\delta\mathbf{B}\|^3).
\label{eq:taylor_expansion}
\end{align}
The Zeeman gradient vector  $\mathbf{S}_1(\mathbf{B}_0)$ [$\mathrm{MHz/mT}$] is the first-order sensitivity and the Hessian matrix $\mathbf{S}_2(\mathbf{B}_0)$ [$\mathrm{MHz/mT^2}$] is the second-order sensitivity \cite{mcauslan2012,g23x-14ks}. These are the derivatives in magnetic-field parameter space:

\begin{align}
\mathbf{S}_1(\mathbf{B}) &= \nabla_{\mathbf{B}} f_{ij}(\mathbf{B}), \label{eq:s1_def} \\
\mathbf{S}_2(\mathbf{B}) &= \nabla_{\mathbf{B}} \nabla_{\mathbf{B}} f_{ij}(\mathbf{B}). \label{eq:s2_def}
\end{align}

The rise over run approach used in finite-difference derivatives requires an arbitrary field increment which gets affected by truncation and cancellation errors. Therefore we use JAX automatic differentiation \cite{jax2018github} to differentiate the implemented transition-frequency function, this approach avoids a finite difference step-size parameter.  (\texttt{jax\_enable\_x64=True}) enables 64-bit floating-point arithmetic for the calculations, for numerical precision and stability. Only where the relevant energy-ordered eigenvalue branches are locally smooth, we apply the derivative treatment. 

 \texttt{jax.grad} provides us the gradient, and the \texttt{jax.jacfwd} renders the Hessian evaluated as its Jacobian \cite{jax2018github}. To minimize iterations and separate argument passing, the gradient is returned as an auxiliary output by \texttt{has\_aux=True}. To ensure that the we are strictly dealing with real values, we remove minor numerical asymmetry arising from floating-point accumulation by explicitly symmetrizing the calculated Hessian:

\begin{equation}
\mathbf{S}_2 \leftarrow \frac{1}{2}\left(\mathbf{S}_2 + \mathbf{S}_2^{\mathsf{T}}\right).
\label{eq:hessian_symmetrize}
\end{equation}
Mathematically, the first-order sensitivity is found such that the Zeeman gradient is zero \cite{mcauslan2012,g23x-14ks}. This is the Zero-First-Order Zeeman (ZEFOZ) operating point $B_0$:

\begin{equation}
\mathbf{S}_1(\mathbf{B}_0) = \mathbf{0}.
\label{eq:zefoz_condition}
\end{equation}
The linear term in Eq.~\eqref{eq:taylor_expansion} vanishes at an exact ZEFOZ coordinate, then the transition frequency detuning is governed by second order sensitivity, which is the Hessian:
\begin{equation}
\delta f_{ij} \simeq \frac{1}{2} \delta\mathbf{B}^{\mathsf{T}} \mathbf{S}_2 \delta\mathbf{B}.
\label{eq:pure_quadratic_detuning}
\end{equation}

$\mathbf{S}_2$ being a real symmetric $3\times3 $ matrix, its spectral decomposition yields 3 real eigenvalues $\lambda_a$ and an orthonormal set of eigenvectors $\mathbf{v_a}$:
\begin{equation}
\mathbf{S}_2 \mathbf{v}_a = \lambda_a \mathbf{v}_a.
\label{eq:hessian_eig}
\end{equation}
The maximum absolute eigenvalue of the Hessian serves as the nonnegative scalar curvature metric $S_{2,\mathrm{max}}$:
\begin{equation}
S_{2,\mathrm{max}} \equiv \max_a |\lambda_a(\mathbf{S}_2)|.
\label{eq:s2max_formal_def}
\end{equation}
Let $a_\star$ be an index satisfying $a_\star \in \operatorname*{arg\,max}_a |\lambda_a|$, and let $\mathbf{v}_{\mathrm{max}} = \mathbf{v}_{a_\star}$. For a displacement
$\delta\mathbf{B}=\delta B_\parallel\mathbf{v}_{\mathrm{max}}$,
the quadratic shift is
\begin{equation}
\delta f_{ij}\simeq
\frac12\lambda_{a_\star}(\delta B_\parallel)^2.
\label{eq:critical_axis_shift}
\end{equation}
 $\mathbf{v}_{\mathrm{max}}$ being the eigenvector for the highest scalar curvature, it acts as a direction of maximum absolute quadratic frequency response. 

Thus, $\mathbf{v}_{\mathrm{max}}$ identifies a direction of maximum absolute quadratic frequency response for a fixed displacement magnitude.It defines an axis of maximum sensitivity, not a specific one-way arrow. It is different from the operating field direction. Its association to dephasing is dependent on the magnetic noise covariance, introduced below. (Note: reported curvature values in tables are converted to $\mathrm{Hz/mT^2}$).

\subsection{Multiseed Newton-Raphson Root Search}
\label{subsec:methods_solver}

We find roots by using a regularized, step-clipped Newton update which are stationary ZEFOZ points that satisfy $\mathbf{S}_1(\mathbf{B}_0) = \mathbf{0}$. The automatically differentiated Hessian is utilized to evaluate and the field vector at iteration $n$ is updated as:
\begin{align}
\mathbf{d}_n &= \left( \mathbf{S}_2(\mathbf{B}_n) + \lambda \mathbf{I}_3 \right)^{-1} \mathbf{S}_1(\mathbf{B}_n), \\
B_{n+1,k} &= B_{n,k} - \operatorname{clip}\left(d_{n,k}, -25~\mathrm{mT}, 25~\mathrm{mT}\right),
\label{eq:newton_step}
\end{align}
where $k \in \{D_1, D_2, b\}$ and $\mathbf{I}_3$ is the three-dimensional identity matrix .  The diagonal shift $\lambda = 10^{-9}~\mathrm{MHz\,mT^{-2}}$ is Levenberg-Marquardt regularization \cite{marquardt1963} introduced to prevent computing the inverse of a singular matrix. The Newton-Raphson calculates the slope of the highly non-linear energy landscape, when the slope is steep, it results into an update which takes a large leap in the magnetic field domain and hence misses the ZEFOZ point entirely. To prevent this, we clip the Cartesian field-step component $k$ to $\pm 25~\mathrm{mT}$.

16 energy levels allows 120 transition pairs $(j>i)$, we create a set of 150 pseudorandom field vectors to eliminate any seed-grid bias, and trajectories are initialized from the same set of these field vectors for each of the 120 transition pairs. Each cartesian component is independent and uniformly drawn from  $[-1000, 1000]~\mathrm{mT}$ (using a fixed random seed of 42), along with the deterministic seed $ (1, 1, 1)~\mathrm{mT}$. This generates a broad range of initial fields.

Iteration halts and trajectories terminate when $\|\mathbf{S}_1\| \le 10^{-10}~\mathrm{MHz/mT}$ which is the internal solver tolerance, or after 100 iterations. Even when a final point satisfies the looser acceptance threshold of  $\|\mathbf{S}_1\| < 10^{-8}~\mathrm{MHz/mT}$ and lies within the imposed search bound $\|\mathbf{B}\| < 2000~\mathrm{mT}$, we retain it for analysis. 

Duplicate points were found since different initial seeds can converge to the same ZEFOZ solution, and were omitted. A rounding convention was adopted in this procedure to group close solutions and pick a representative value for each group.

\subsection{Lattice-Informed \texorpdfstring{$^{89}\mathrm{Y}$}{89Y} Magnetic Bath}
\label{subsec:methods_bath}

$^{167}\mathrm{Er}^{3+}$ ion experiences a local spin-bath magnetic environment of the $\mathrm{Y}_2\mathrm{SiO}_5$ host crystal lattice. 
The symmetrized $\mathrm{Y}_2\mathrm{SiO}_5$ crystal structure was obtained from the Materials Project database (mp-3520)~\cite{materialsproject_yso}. The CIF was processed using pymatgen~\cite{pymatgen2013} to extract the lattice vectors and crystallographic Y-site coordinates. To model this spin-bath, the monoclinic  $\mathrm{Y}_2\mathrm{SiO}_5$ ($C2/c$, space group 15) is expanded across translation from $-3$ to $+3$ conventional cells along each lattice direction. $\mathrm{Er}^{3+}$ is doped at Site 2 substitutional $\mathrm{Y}$ position \cite{materialsproject_yso}. The site is identified as Site 2 via its 7-fold oxygen coordination \cite{sun2008,g23x-14ks}. The coordinate where the dopant sits is excluded from the host-spin summation and the neighboring nuclear coordinates are taken from the unrelaxed host structure.

The dipolar field coupling tensor $\mathbf{D}_k$ expresses the coupling between dopant origin and the $k$-th $^{89}\mathrm{Y}$ host nucleus ($I=1/2$, $100\%$ natural abundance)\cite{nistY} at displacement $\mathbf{r}_k$:
\begin{equation}
\mathbf{D}_k = \frac{\mu_0 \hbar |\gamma_Y|}{4\pi r_k^3} \left( 3 \hat{\mathbf{r}}_k \hat{\mathbf{r}}_k^{\mathsf{T}} - \mathbf{I}_3 \right).
\label{eq:dipolar_tensor}
\end{equation}
 $\mathbf{D}_k$ has units of \textbf{tesla (T)} when distances are in SI units and $|\gamma_Y| = 2\pi \times 2.0949~\mathrm{MHz/T}$ is the gyromagnetic ratio in angular-frequency units. The field contribution $\delta\mathbf{B}_k = \mathbf{D}_k \mathbf{I}_k$ is produce by the dimensionless nuclear-spin operator $\mathbf{I}_k$ \cite{abragam1961,ulanowski2025}. 

Under High-Temperature approximation condition the thermal energy $k_B T$ is considerably higher than the Zeeman splitting energy  $\hbar|\gamma_Y|B $, due to this the spins are vastly disordered, unpolarized state which points to random equally likely directions \cite{abragam1961,vanvleck1948}. So, the equal-time spin second moment is approximated by $\langle I_{k,\alpha} I_{j,\beta} \rangle = \frac{1}{4}\delta_{kj}\delta_{\alpha\beta}$. The raw equal-time covariance of the local magnetic field at the dopant is:
\begin{equation}
\boldsymbol{\Sigma}_{B,\mathrm{raw}} = \frac{1}{4} \sum_{k=1}^N \mathbf{D}_k \mathbf{D}_k^{\mathsf{T}}.
\label{eq:covariance_sum}
\end{equation}
The average directional noise amplitude or the equivalent isotropic raw noise amplitude is derived from the trace (Tr) of the covariance matrix as $\sigma_{\mathrm{iso,raw}} = [\operatorname{Tr}(\boldsymbol{\Sigma}_{B,\mathrm{raw}})/3]^{1/2}$. As we gradually increased the bath size, the radial inclusion cutoff from $10$ to $18$~\AA, we observed an increase in the number of $^{89}$Y nuclei from $83$ to $443$, but $\sigma_{\mathrm{iso,raw}}$ grew from $5.703$ to $5.753~\mu\mathrm{T}$. This highlights the dominant influence of nearby nuclei: individual dipolar fields scale as  $r^{-3}$, and therefore their covariance contributions scale as $r^{-6}$. The proxy inverse RMS-coupling correlation-time reduces from $13.481$ to $12.207~\mathrm{ms}$ as more Y--Y couplings are included and finite-cluster boundary effects are reduced. The important observation which helped designate $15$~\AA\ as the cutoff radius, was the negligible increment in the raw RMS field by only $0.13\%$ as we extend the cutoff from $15$ to $18$~\AA, while the correlation-time proxy by approximately $1.9\%$, despite increasing the bath population from  $267$ to $443$. Hence the bath descriptors converged.

\subsection{Correlation-Time Proxy and Covariance Normalization}
\label{subsec:methods_transform}

In order to have a metric for the dynamical fluctuations of the bath, the homonuclear $^{89}\mathrm{Y}\text{--}^{89}\mathrm{Y}$ secular dipolar coupling statistics is used to estimate a correlation-time proxy. The secular coupling magnitude is
\begin{equation}
b_{ij} = \frac{\mu_0\hbar|\gamma_Y|^2}{4\pi r_{ij}^{3}} \left|1-3\cos^2\theta_{ij}\right|,
\label{eq:homonuclear_rate}
\end{equation}
where $\theta_{ij}$ is the angle between the internuclear displacement and the provisional quantization axis (the crystallographic $b$-axis) \cite{abragam1961,vanvleck1948}. The couplings are evaluated in angular-frequency units.

For each bath spin $i$, we calculate the local coupling summary:
\begin{equation}
b_{i,\mathrm{RMS}} = \left( \sum_{j \in \mathcal{N}_i} b_{ij}^2 \right)^{1/2},
\label{eq:local_rms_coupling}
\end{equation}
where $\mathcal{N}_i$ contains the other retained bath spins within the specified Y--Y pair-distance cutoff ($15.0$~\AA). The central correlation-time proxy is then defined as:
\begin{equation}
\tau_{\mathrm{RMS}} = \left[ \frac{1}{N} \sum_i b_{i,\mathrm{RMS}} \right]^{-1}.
\label{eq:tau_rms}
\end{equation}
This gives $\tau_{\mathrm{RMS}} = 12.44~\mathrm{ms}$, this quantity is a coupling based-dynamical proxy and is not a correlation time extracted from a simulated or measured field-autocorrelation function. Alternatively, we also calculate the inverse-sum  ($\tau_{\mathrm{sum}} = 2.70~\mathrm{ms}$) and inverse-maximum ($\tau_{\mathrm{max}} = 21.12~\mathrm{ms}$) coupling proxies merely as a model estimator comparison and not as a rigorous physical bound on the bath correlation time. In our coherence model introduced below, we adopt  $\tau_c=\tau_{\mathrm{RMS}}$ as the correlation-time parameter for the evaluated transitions because where $\tau_{\mathrm{sum}}$ and $\tau_{\mathrm{max}}$ adds magnitudes linearly or retains the strongest pair, $\tau_\mathrm{RMS}$ includes all local couplings through their second moment.  

The microscopic lattice sum helps us obtain the field-noise covariance and is used directly without any phenomenological amplitude rescaling. Its equivalent single-axis RMS scale is 

\begin{equation}
\sigma_{\mathrm{iso}}
=
\sqrt{\frac{\operatorname{Tr}
(\boldsymbol{\Sigma}_{B,\mathrm{raw}})}{3}}.
\label{eq:equivalent_rms}
\end{equation}
$\sigma_{\mathrm{iso}}=5.745~\mu\mathrm{T}$ for the selected lattice cutoff. This is an equivalent per-component RMS, not the most probable magnitude of the three-dimensional field. The overall noise amplitude and its spatial anisotropy, they both follow from the adopted lattice bath model. $\Sigma_{B,raw}$ trace is not scaled, to do so, the covariance is transformed into the Hamiltonian frame as

\begin{equation}
\boldsymbol{\Sigma}_{B,(D_1,D_2,b)}
=
\mathbf{R}\boldsymbol{\Sigma}_{B,\mathrm{raw}}
\mathbf{R}^{\mathsf{T}}.
\label{eq:sigma_rotation}
\end{equation}

We construct an orthonormal reference basis in CIF coordinates, define  $\hat{\mathbf{e}}_b=\mathbf{b}_{\mathrm{CIF}}/ |\mathbf{b}_{\mathrm{CIF}}|$ and $\hat{\mathbf{e}}_a$ as the normalized projection of $\mathbf{a}_{\mathrm{CIF}}$ perpendicular to $\mathbf{b}_{\mathrm{CIF}}$, and set $\hat{\mathbf{e}}_p=\hat{\mathbf{e}}_b\times\hat{\mathbf{e}}_a$.

The CIF and spectroscopic reference use different cell descriptions. We adopt the correspondence
\begin{align}
\mathbf{a}_{\mathrm{sp}} &= \mathbf{c}_{\mathrm{CIF}},\\
\mathbf{b}_{\mathrm{sp}} &= \mathbf{b}_{\mathrm{CIF}},\\
\mathbf{c}_{\mathrm{sp}} &=
-(\mathbf{a}_{\mathrm{CIF}}+\mathbf{c}_{\mathrm{CIF}}).
\end{align}
This gives transformed lattice parameters $(a_{\mathrm{sp}},b_{\mathrm{sp}},c_{\mathrm{sp}}) =(10.421, 6.743, 12.518)$~\AA\ and $\beta_{\mathrm{sp}}=102.603^\circ$.
The separation between $D_1$ and $\mathbf{a}_{\mathrm{sp}}$ is reported as $78.7^\circ$~\cite{sun2008}, using this value for the selected orientation branch gives
\begin{equation}
\theta_{D_1}
=180^\circ-\beta_{\mathrm{CIF}}-78.7^\circ
=-20.887785^\circ
\pmod{180^\circ},
\end{equation}
where $\beta_{\mathrm{CIF}}=122.18778465^\circ$.
We therefore use $\theta_{D_1}=-20.9^\circ$.

This construction enables us to transfer the experimental optical-axis orientation to our evaluated lattice without independently calculating its dielectric principle axes.

The dielectric-axis unit vectors are
\begin{align}
\hat{\mathbf{e}}_{D_1}
&=\cos\theta_{D_1}\,\hat{\mathbf{e}}_a
+\sin\theta_{D_1}\,\hat{\mathbf{e}}_p,
\label{eq:d1_rot}\\
\hat{\mathbf{e}}_{D_2}
&=-\sin\theta_{D_1}\,\hat{\mathbf{e}}_a
+\cos\theta_{D_1}\,\hat{\mathbf{e}}_p.
\label{eq:d2_rot}
\end{align}
The rows of $\mathbf{R}$ are:
\begin{equation}
\mathbf{R} = 
\begin{pmatrix}
\text{---} & \hat{\mathbf{e}}_{D_1}^{\mathsf{T}} & \text{---} \\
\text{---} & \hat{\mathbf{e}}_{D_2}^{\mathsf{T}} & \text{---} \\
\text{---} & \hat{\mathbf{e}}_b^{\mathsf{T}} & \text{---}
\end{pmatrix}.
\end{equation}

\subsection{Full-Hessian Quadratic Frequency-Noise Model}
\label{subsec:full-hessian-kubo}

At a ZEFOZ point, the prominent transition-frequency response is quadratic as the first-order magnetic-field sensitivity vanishes \cite{mcauslan2012,g23x-14ks}:
\begin{equation}
\delta f \simeq \frac{1}{2}\,\delta\mathbf{B}^{\mathsf T}\mathbf{S}_2\,\delta\mathbf{B},
\label{eq:quadratic_response}
\end{equation}
where \(\mathbf{S}_2\) is the symmetric frequency Hessian and \(\delta\mathbf{B}\) is a magnetic-field fluctuation about the ZEFOZ point.

For zero-mean Gaussian magnetic-field fluctuations with equal-time covariance
\begin{equation}
\Sigma_{B,\alpha\beta}
=
\left\langle
\delta B_\alpha\delta B_\beta
\right\rangle,
\end{equation}
the variance of the centered quadratic frequency fluctuation is \cite{isserlis1918}
\begin{equation}
\operatorname{Var}(\delta f)
=
\frac{1}{2}
\operatorname{Tr}
\left[
\mathbf{S}_2\boldsymbol{\Sigma}_B
\mathbf{S}_2\boldsymbol{\Sigma}_B
\right].
\label{eq:freq_variance}
\end{equation}
We convert the Hessian from \(\mathrm{MHz/mT^2}\) to \(\mathrm{Hz/mT^2}\) before this evaluation. The corresponding angular-frequency amplitude is
\begin{equation}
\Delta_\omega
=
2\pi
\sqrt{\operatorname{Var}(\delta f)}.
\label{eq:angular_freq_amplitude}
\end{equation}

For an exponential Gaussian field-correlation model,
\begin{equation}
\mathbf{C}_B(t)
=
\boldsymbol{\Sigma}_B e^{-|t|/\tau_c},
\label{eq:exponential_correlation}
\end{equation}
the centered quadratic frequency-noise correlation decays as \(e^{-2|t|/\tau_c}\) \cite{kubo1962,anderson1954}. The effective quadratic-noise correlation time is therefore
\begin{equation}
\tau_{\mathrm{quad}}
=
\frac{\tau_c}{2}.
\label{eq:quadratic_correlation_time}
\end{equation}
The dimensionless Kubo parameter is
\begin{equation}
\kappa
=
\Delta_\omega\tau_{\mathrm{quad}}.
\label{eq:kubo_kappa}
\end{equation}

The magnetic contribution to the coherence function is evaluated in the second-cumulant approximation \cite{kubo1962,anderson1954} as
\begin{equation}
\chi_{\mathrm{mag}}(t)
=
\exp[-K(t)],
\label{eq:coherence_function}
\end{equation}
with
\begin{equation}
K(t)=\left(\Delta_\omega\tau_{\mathrm{quad}}\right)^2\left[e^{-t/\tau_{\mathrm{quad}}}+\frac{t}{\tau_{\mathrm{quad}}}-1\right].
\label{eq:kubo_integral}
\end{equation}
The modeled Y-bath-induced quadratic magnetic pure-dephasing time \(T_{\phi,Y}^{(2)}\) is defined by the condition
\begin{equation}
K(T_{\phi,Y}^{(2)})=1.
\end{equation}
In the fast-modulation limit, \(\kappa\ll1\), this reduces to
\begin{equation}
\frac{1}{T_{\phi,Y}^{(2)}} \simeq \Delta_\omega^2 \tau_{\mathrm{quad}}
\end{equation}
\(\Delta_\omega\) is acquired from full Hessian and the full covariance tensor, all magnetic-field components and their correlations are included in this estimate

It is meaningful to demand physically interpretable comparison with a simplified one-axis treatment, for that we also evaluate the field-noise variance projected onto the critical hessian axis,
\begin{equation}
\sigma_\parallel^2
=
\mathbf{v}_{\mathrm{max}}^{\mathsf T}
\boldsymbol{\Sigma}_B
\mathbf{v}_{\mathrm{max}},
\label{eq:critical_axis_diagnostic}
\end{equation}
where \(\mathbf{v}_{\mathrm{max}}\) is the eigenvector associated with the largest-magnitude eigenvalue of \(\mathbf{S}_2\). In presence of a fluctuating field that points directly along the \(\mathbf{v}_{\mathrm{max}}\) direction, will cause the induction of the maximum frequency shift possible for a given fluctuation magnitude.  This projection measures the magnetic-field noise along the direction to which the transition is most quadratically sensitive.

It is retained as a worst-case diagnostic quantity only. The primary coherence estimates use the full Hessian  \(\mathbf{S}_2\) and the full covariance tensor \(\boldsymbol{\Sigma}_B\). \(\sigma_\parallel^2\) does not take account for cross-terms between field components or the contribution of other Hessian eigendirections.

The equivalent per-component RMS field,
\begin{equation}
\sigma_{\mathrm{iso}}=\sqrt{\frac{\operatorname{Tr}\boldsymbol{\Sigma}_B}{3}},
\label{eq:equivalent_rms_2}
\end{equation}
is likewise reported only as a scalar summary of the covariance magnitude. It is not substituted for the full anisotropic covariance tensor in the coherence calculation.

\subsection{Unpolarized Erbium Ensemble Second Moment}
\label{subsec:er-second-moment}

We probe the magnetic-field second moment generated by a dilute, fully unpolarized $\mathrm{Er}^{3+}$ electronic-spin ensemble using the crystallographic lattice. $\mathrm{Er}$ substitution is modeled as an independent stochastic occupation of native $\mathrm{Y}$ sites, deliberately excluding the central reference ion from the summation. This calculation characterizes the occupation-averaged dipolar field covariance at the central ion- it does not calculate a dynamic electronic exchange-coupling rate\cite{bottger2006,g23x-14ks}.

A 15~\AA cutoff securely captures the dense Y-bath. Erbium, however, is highly dilute; the low Er occupation probability motivates extending the Er lattice sum to larger distances to examine convergence\cite{bottger2006,g23x-14ks}. We therefore evaluate cutoffs up to 150~\AA, where the expanded volume incorporates both the Site-1 and Site-2 electronic $\mathbf{g}$ tensors alongside their symmetry-related magnetic orientations \cite{sun2008,wang2023}. The respective moment tensors are assigned directly onto the substitutional Y positions using the crystallographic symmetry operations, without imposing an electronic resonance or same-subclass selection criterion. 

For independent lattice occupations $n_j\in\{0,1\}$ with occupation probability $c$ ($c=10^{-5}$ at $10$ atomic ppm), the unpolarized Er magnetic-field covariance within a radial cutoff $R$ is \cite{vanvleck1948,abragam1961}
\begin{equation}
\boldsymbol{\Sigma}_{B,\mathrm{Er}}(R)=c\sum_{j:r_j\leq R}\mathbf{D}_j\mathbf{M}_j\mathbf{D}_j^{\mathsf T},
\label{eq:er_field_covariance}
\end{equation}

where $\mathbf{D}_j=\mu_0\mu_B(3\hat{\mathbf r}_j\hat{\mathbf r}_j^{\mathsf T}-\mathbf I_3)/(4\pi r_j^3)$ is the dipolar-field tensor. When evaluating an unpolarized effective electronic spin $S=1/2$, the dimensionless magnetic-moment second-moment tensor resolves to $\mathbf{g}^{\mathsf T}\mathbf{g}/4$ within the adopted optical frame. To synthesize $\mathbf{M}_j$, the tensor is transformed into the CIF Cartesian frame and rotated, obeying the crystallographic symmetry operation associated with each distinct substitutional site. The factor $c$ accounts for the independent statistical probability of Er lattice occupation, excluding the central substitutional position from the summation.

Extracting the trace yields the equivalent per-component RMS field amplitude, obtained as $\sigma_{\mathrm{iso,Er}}=[\operatorname{Tr}(\boldsymbol{\Sigma}_{B,\mathrm{Er}})/3]^{1/2}$. Its dependence on the spatial cutoff characterizes the convergence of the occupation-averaged electronic field second moment. Yet, this limit faces a profound statistical vulnerability in a scenario where rare, closely spaced Er substitutions may contribute disproportionately to this quantity by inflating the summation. Consequently, the resulting RMS field does not necessarily represent the magnetic environment of a typical ion and this calculation does not determine electronic flip-flop rates, temporal correlations, or spectral-diffusion-induced dephasing, so, Er covariance is therefore kept separate from the primary Y-only coherence model. At an intended operational baseline of approximately $10~\mathrm{mK}$, the substantial magnetic bias fields at the identified ZEFOZ points are expected to produce strong polarization of the surrounding Er electronic-spin ensemble, suppressing resonant Er-Er flip-flop dynamics \cite{bottger2006,budoyo2018,ulanowski2025}. This physically justifies neglecting dynamic Er- induced dephasing in the leading-order low-temperature approximation. We retain the unpolarized Er field second moment, not as a dynamical parameter but strictly as a separate,  geometrically determined structural characterization of the dilute dopant ensemble.

\subsection{Microwave Coupling and Transition Comparison}
\label{subsec:microwave-pareto}

Microwave addressability is a property of a transition that determines how strongly it can be driven for coherent spin control. For a given microwave magnetic-field amplitude, it determines how quickly we can flip or manipulate the stored spin state using a microwave pulse. We calculate the electronic magnetic-dipolar coupling using the drive Hamiltonian $\mathcal{H}_{\mathrm{drive}}(t)$ \cite{rakonjac2020,wang2023} in frequency units,

\begin{equation}
\mathcal{H}_{\mathrm{drive}}(t) \equiv \frac{H_{\mathrm{drive}}(t)}{h} = \frac{\mu_B}{h} \sum_{k \in \{D_1, D_2, b\}} B_{1,k}(t) (\mathbf{g}_0 \mathbf{S})_k,
\label{eq:mw_drive}
\end{equation}

Here $\mathbf{S}$ is the dimensionless effective electron spin operator. The microwave magnetic field couples to the electronic magnetic moment and can drive transitions between the coupled electron-nuclear states eigenstates. The vector $\mathbf{B}_1(t)$ is the applied oscillating AC magnetic field from the microwave pulse resolved along the three axes of the optical coordinate frame \((D_1, D_2, b)\)

For a transition between eigenstates $\lvert i \rangle$ and $\lvert j \rangle$, we define the transition vector
\begin{equation}
m_k = \left\langle j \left| \frac{\partial \mathcal{H}_{\mathrm{drive}}}{\partial B_{1,k}} \right| i \right\rangle,
\label{eq:transition_vector}
\end{equation} expressed in $\mathrm{MHz/mT}$. The magnitude gives the coupling in \(\mathrm{MHz}\) for a field amplitude of \(1~\mathrm{mT}\) along that axis. 

For a real, unit-normalized linear polarization vector $\hat{\mathbf{p}}$, the coupling strength is
\begin{equation}
M(\hat{\mathbf{p}}) = \left| \hat{\mathbf{p}}^{\mathsf T} \mathbf{m} \right|.
\end{equation}which specifies the spatial alignment of the linearly polarized microwave relative to the crystal's optical coordinate frame. The dot product $\left|\hat{\mathbf{p}}^{\mathsf T} \mathbf{m} \right|$ calculates the effective driving strength for that specific local field polarization.
The optimal coupling over real linear polarizations is
\begin{equation}
M_{\mathrm{opt}} = \max_{\substack{\hat{\mathbf{p}}\in\mathbb{R}^3\\ \|\hat{\mathbf{p}}\|=1}} \left| \hat{\mathbf{p}}^{\mathsf T} \mathbf{m} \right|.
\label{eq:m_opt}
\end{equation}This maximization identifies the linear microwave-field polarization that drives the transition with good strength for a fixed field amplitude. We form the real symmetric matrix  \(\mathbf{G}=\operatorname{Re}(\mathbf{m}\mathbf{m}^{\dagger})\). The square root of its largest value gives \(M_{\mathrm{opt}}\) 

For a resonant drive $\mathbf{B}_1(t)=B_1\hat{\mathbf{p}}_{\mathrm{opt}} \cos(2\pi f_{ij}t)$, where $B_1$ is the peak field amplitude and $\hat{\mathbf{p}}_{\mathrm{opt}}$ is an optimizing linear polarization, the rotating-wave approximation \cite{abragam1961} gives
\begin{equation}
\frac{f_R}{B_1} = 10^6 M_{\mathrm{opt}}
\quad \mathrm{Hz/mT},
\label{eq:rabi_per_field}
\end{equation}
when $M_{\mathrm{opt}}$ is expressed in
$\mathrm{MHz/mT}$. $f_R$ is the Rabi frequency.

The evaluated ZEFOZ croots are screened on the basis of experimental viability using the adopted operating-point critteria:
\begin{equation}
\begin{aligned}
\|\mathbf{B}_0\| &\ge 1.0~\mathrm{mT}, \qquad 0<f_{ij}\le 2000~\mathrm{MHz},\\
\kappa &\le 10^{-2}, \qquad T_{\phi,Y}^{(2)}\ge 1000~\mathrm{s},\\
M_{\mathrm{opt}} &\ge 10^{-4}~\mathrm{MHz/mT}.
\end{aligned}
\label{eq:table_fast_condition}
\end{equation}
The thresholds we adopted for screening are not universal requirements for experimental viability. A transition that fails these condition is not necessarily experimentally unusable, subject to the the experimental setup. These conditions make the experiment possible and physically realizable to a greater extent. We restrict the magnetic field for minimum bias, where stray-field compensation would require separate consideration. The microwave frequency is bounded to ensure the transition is non-degenerate and practically addressable. Deep fast modulation ensures that quadratic frequency noise decorrelates safely before appreciable coherence is lost, the retained lifetimes exhibit motional-narrowing, approximately exponential decay and closely agree with $\frac{1}{T_{\phi,Y}^{(2)}} \simeq \Delta_\omega^2 \tau_{\mathrm{quad}}$. The other conditions prioritizes substantial headroom before additional experimental limitation are considered. 
These times are modeled Y-bath magnetic pure-dephasing estimates acquired by solving \(K(T_{\phi,Y}^{(2)})=1\) using the full-Hessian model; Er-induced dephasing and population relaxation is excluded. Nine distinct transitions pair are found eligible after screening and are included in the table. 

Then to identify a coherence-control trade-off, we evaluate Pareto membership among these nine representatives where a candidate is dominated if another has both  an equal or longer modeled pure-dephasing time and an equal or stronger microwave coupling, at least one strict improvement. Four representatives emerge non-dominated and form the Pareto frontier. It is an annotation rather than an additional inclusion filter. 

\subsection{Hamiltonian-Uncertainty Sampling and Numerical Recalibration}
\label{subsec:uncertainty-recalibration}

There can be a difference between nominal literature parameters and the effective spin Hamiltonian of a physical sample \cite{welinski2017,galland2020}, this can shift the position of a ZEFOZ point. To see the effect of such parameter deviations and evaluate the sensitivity of the identified operating point, we implement an agnostic Hamiltonian-parameter perturbation test. 

We generate $N = 500$ sampled Hamiltonians by applying independent relative perturbations to the upper-triangular elements of the nominal tensors for each tested candidate and then the perturbed elements are copied to their symmetric counterparts:
\begin{equation}
T_{ij}^{(s)} = (T_0)_{ij} \left(1 + \delta_{ij}^{(s)} \right), \qquad T \in \{A, Q, g\}.
\end{equation}
Here, \(\delta_{ij}^{(s)}\sim\mathcal{N}(0,\sigma_T^2)\) for \(i\le j\), and \(T_{ji}^{(s)}=T_{ij}^{(s)}\). We use \(\sigma_A=\sigma_g=0.01\) and \(\sigma_Q=0.02\). These perturbations are multiplicative, so nominally zero tensor elements remain zero. This generates a sensitivity ensemble with independently sampled upper-triangular tensor elements rather than a experimentally measured uncertainty distribution inferred from correlated tensor fit or measured strain model. 

For each sampled Hamiltonian, we evaluate two distinct operational scenarios:
\begin{enumerate}
    \item \textbf{Open-Loop Operation:} The applied field is \(\mathbf{B}_{\mathrm{open}}^{(s)}=\mathbf{B}_0+\delta\mathbf{B}_{\mathrm{set}}^{(s)}\),  where the field-setting error has independent zero-mean Gaussian components with a standard deviation of \(0.10~\mathrm{mT}\) along each axis of the \((D_1,D_2,b)\) frame. Generally, as a consequence of Hamiltonian perturbations and field-setting errors, residual first-order sensitivity \(\mathbf{S}_1\neq\mathbf{0}\) is produced and the resulting coherence estimate includes the causal linear magnetic-noise contribution with the prevailing full-Hessian quadratic contribution. 
    \item \textbf{Numerical Recalibration:} A regularized, step-limited Newton iteration is used starting from \(\mathbf{B}_0\) with the full Hessian of the perturbed Hamiltonian to  identify a recalibrated operating point  \(\mathbf{B}_{\mathrm{recal}}^{(s)}\). Convergence requires \(\|\mathbf{S}_1(\mathbf{B}_{\mathrm{recal}}^{(s)};\mathbf{A}^{(s)},\mathbf{Q}^{(s)},\mathbf{g}^{(s)})\|<10^{-8}~\mathrm{MHz/mT}\). 
We limit the iteration to 20 steps and constrain the Euclidean norm of each Newton field update to $10~\mathrm{mT}$.
 
\end{enumerate}

Open-loop and recalibrated coherence estimates are carried out using the fast-modulation linear and full-Hessian quadratic magnetic-dephasing rates, whereas the Y-field covariance tensor and its adopted correlation-time proxy are held the same throughout this test. 

Post successful recalibration, the samples undergo vector-field relocation magnitude evaluation: 
\begin{equation}
\Delta B^{(s)} = \left\| \mathbf{B}_{\mathrm{recal}}^{(s)} - \mathbf{B}_0 \right\|.
\label{eq:field_relocation}
\end{equation}
To characterize the robustness of a transition, its numerical recovery fraction is analyzed which is defined as the number of successfully recaibrated samples divided by 500, and by the distributions of the recalibrated pure-dephasing time and field relocation among the successful samples. Pure-dephasing-time distributions are summarized by the median- describing the central sampled coherence, and the \(15.9\)th-\(84.1\)st percentile interval- containing approximately 68.2\% of the distribution without assuming symmetry or Gaussianity. For field relocation, the median describes the central adjustment and the \(84.1\)st percentile indicates a larger adjustment enough for approximately 84.1\% of successfully recalibrated samples.

\section{ZEFOZ Operating-Point Landscape}
\label{sec:landscape}

\subsection{Located Stationary Points and Search Coverage}
\label{subsec:located-roots}

The multiseed Newton-Raphson search obtained 344 stationary ZEFOZ point candidates satisfying the $\|\mathbf{S}_1\|<10^{-8}~\mathrm{MHz/mT}$ and $\|\mathbf{B}_0\|<2000~\mathrm{mT}$ criteria, after grouping solutions by transition pair and magnetic field components rounded to $0.1~\mathrm{mT}$.  The solver tolerance threshold is of $10^{-10}~\mathrm{MHz/mT}$ and for each transition pair, the iteration is initialized from 150 pseudorandom field vectors and one additional fixed vector. The random initialization avoids regular-grid bias. Multiple seeds can converge to to the same point, and we cannot establish exhaustiveness as some solutions may have regions of convergence that are not reached within the numerical iteration limits.

The search recovers the Site-2 $\epsilon_{14}\leftrightarrow\epsilon_{15}$ ZEFOZ operating point reported by Matsuura et al., obtaining a field magnitude of $633.527~\mathrm{mT}$, lying close to their reported value of $633.52~\mathrm{mT}$~\cite{g23x-14ks}. This agreement gives a consistency check for the effective-Hamiltonian implementation and root-finding procedure. 

\subsection{Curvature Hierarchy and Local Quadratic Response}
\label{subsec:curvature-hierarchy}

The roots found in the search are ranked by $S_{2,\mathrm{max}}=\max_a|\lambda_a(\mathbf{S}_2)|$, where $\lambda_a$ are the eigenvalues of the frequency Hessian.  Table~\ref{tab:curvature_hierarchy} highlights the five lowest-curvature operating points.

\begin{table*}[t]
\centering
\caption{The five lowest-curvature Site-2 ZEFOZ operating points found in the search, ranked by $S_{2,\mathrm{max}}=\max_a|\lambda_a(\mathbf{S}_2)|$. Field-orientation angles $(\theta,\phi)$ follow the convention implemented in Cell 0. Critical-axis vectors are expressed in the $(D_1,D_2,b)$ frame; their overall signs are arbitrary.}
\label{tab:curvature_hierarchy}
\begin{tabular}{l c c c c c}
\hline\hline
Transition & $\|\mathbf{B}_0\|$ (mT) & $\theta$ ($^\circ$) & $\phi$ ($^\circ$) & $S_{2,\mathrm{max}}$ (Hz/mT$^2$) & Critical Axis $\mathbf{v}_{\mathrm{max}}$ \\
\hline
$\epsilon_{14} \leftrightarrow \epsilon_{15}$ & 633.527 & 37.54 & -10.94 & 165.241 & $[-0.7225, +0.3760, +0.5802]$ \\
$\epsilon_{6} \leftrightarrow \epsilon_{7}$ & 668.529 & 38.17 & -12.60 & 173.068 & $[-0.6944, +0.3724, +0.6158]$ \\
$\epsilon_{13} \leftrightarrow \epsilon_{14}$ & 413.741 & 40.22 & -11.85 & 282.525 & $[-0.6646, +0.4353, +0.6072]$ \\
$\epsilon_{13} \leftrightarrow \epsilon_{15}$ & 510.470 & 38.88 & -11.42 & 432.162 & $[-0.6965, +0.4013, +0.5948]$ \\
$\epsilon_{12} \leftrightarrow \epsilon_{13}$ & 283.919 & 42.24 & -12.59 & 503.778 & $[-0.6041, +0.5019, +0.6189]$ \\
\hline\hline
\end{tabular}
\end{table*}

The $\epsilon_{14}\leftrightarrow\epsilon_{15}$ transition operating point has the least maximum absolute directional curvature over all unit field-displacement directions among the located roots, that is, this point has the smallest worst-direction quadratic frequency sensitivity; it does not necessarily have the smallest curvature along every direction. However, \(S_{2,\mathrm{max}}\) involves only the local transition-frequency geometry. Subsequently, Y-bath pure-dephasing estimates will incorporate lattice-derived field-noise covariance and an exponential temporal-correlation model with a correlation time estimated from Y-Y dipolar coupling statistics. 

The critical axis vector $\mathbf{v}_{\mathrm{max}}$  is a metric which identifies the direction of highest absolute quadratic frequency response for a fix field-displacement magnitude. The covariance projection, $\mathbf{v}_{\mathrm{max}}^{\mathsf T}\boldsymbol{\Sigma}_B\mathbf{v}_{\mathrm{max}}$, is a diagnostic tool providing the magnetic-field noise variance along this direction. The coherence calculation can utilize this metric for an estimated dominated single direction dephasing, nevertheless, we retain the full Hessian and covariance tensor for coherence calculation.

The critical-axis vector $\mathbf{v}_{\mathrm{max}}$ identifies the direction of maximum absolute quadratic frequency response for a fixed field-displacement magnitude. The covariance projection $\mathbf{v}_{\mathrm{max}}^{\mathsf T}\boldsymbol{\Sigma}_B\mathbf{v}_{\mathrm{max}}$ measures the magnetic-field noise variance along this direction and is retained as a directional diagnostic. The Y-only coherence calculation instead incorporates the full Hessian and covariance tensor.

At a numerically converged ZEFOZ point, the first-order frequency sensitivity the first-order sensitivity lies below the specified numerical threshold, and the local frequency response is approximated by its quadratic hessian contribution. Along the critical axis, $\delta\mathbf{B}=\delta B_\parallel\mathbf{v}_{\mathrm{max}}$, this response is
\begin{equation}
\delta f \simeq \frac{1}{2}\lambda_{\mathrm{crit}}(\delta B_\parallel)^2,
\qquad
|\lambda_{\mathrm{crit}}|=S_{2,\mathrm{max}},
\end{equation}
where $\lambda_{\mathrm{crit}}$ is the signed Hessian eigenvalue associated with $\mathbf{v}_{\mathrm{max}}$.

\begin{figure*}[t]
\centering
\includegraphics[width=0.96\textwidth]{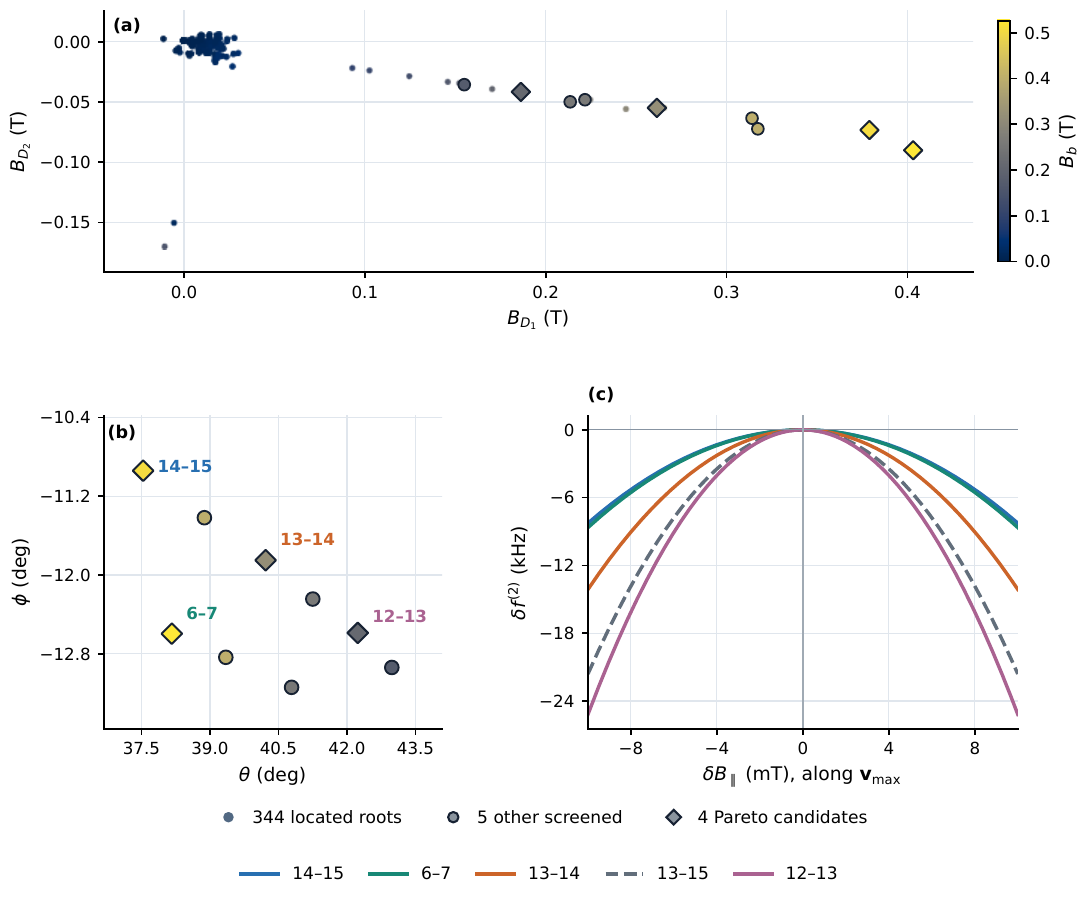}
\caption{Landscape of the located Site-2 ZEFOZ operating points. (a) Projection of the 344 numerical roots onto the $(B_{D_1},B_{D_2})$ plane, with marker colors representing $B_b$. Filled circles denote the five additional screened candidates, while diamonds identify the four Pareto-optimal transitions. (b) Angular coordinates of the nine screened operating points in the $(D_1,D_2,b)$ frame. (c) Signed quadratic frequency response along the critical Hessian eigenvector for the five lowest-curvature transitions.}
\label{fig:zefoz-landscape}
\end{figure*}

\section{Lattice Bath Statistics and Conditional Coherence}
\label{sec:bath-coherence}

\subsection{\texorpdfstring{$^{89}\mathrm{Y}$}{89Y} Bath Convergence, Anisotropy, and Full-Hessian Coherence}
\label{subsec:y-only-coherence}

The central $^{167}\mathrm{Er}^{3+}$ ion experiences a local magnetic environment created by the fluctuating magnetic field due to surrounding $^{89}\mathrm{Y}$ nuclei \cite{ulanowski2025,bottger2006}. Owing to its nuclear spin $I=1/2$, the Y nucleus generates a magnetic dipolar field whose magnitude depends on its distance from the Er ion and whose direction depends on their relative position. Its contribution to the coherence calculation is characterized through a real-space magnetic-dipole lattice sum. We identify the sevenfold oxygen-coordinated Y environment- \texttt{CIF\_Y0}, corresponding to spectroscopic Site 2 in the adopted structure \cite{materialsproject_yso,sun2008}. For Site-2 $\mathrm{Er}^{3+}$ substitution, the equivalent isotropic RMS field amplitude  ($\sigma_{\mathrm{iso}}$) converges tightly as the radial inclusion cutoff is expanded. Table~\ref{tab:lattice_convergence} summarizes this geometric convergence.

\begin{table}[h]
\centering
\caption{Geometric convergence of the unpolarized $^{89}\mathrm{Y}$ field-noise RMS scale and correlation-time proxies for the Site-2 ($\mathrm{CIF\_Y0}$) orbit as a function of the spatial inclusion cutoff.}
\label{tab:lattice_convergence}
\begin{tabular}{c c c c c}
\hline\hline
Cutoff (\AA) & $N_{\mathrm{bath}}$ & $\sigma_{\mathrm{iso}}$ ($\mu$T) & $\tau_{\mathrm{sum}}$ (ms) & $\tau_{\mathrm{RMS}}$ (ms) \\
\hline
10.0 & 83  & 5.703 & 3.854 & 13.481 \\
12.0 & 129 & 5.725 & 3.305 & 13.103 \\
15.0 & 267 & 5.745 & 2.696 & 12.444 \\
18.0 & 443 & 5.753 & 2.370 & 12.207 \\
\hline\hline
\end{tabular}
\end{table}

\begin{figure}[t]
\centering
\includegraphics[width=\columnwidth]{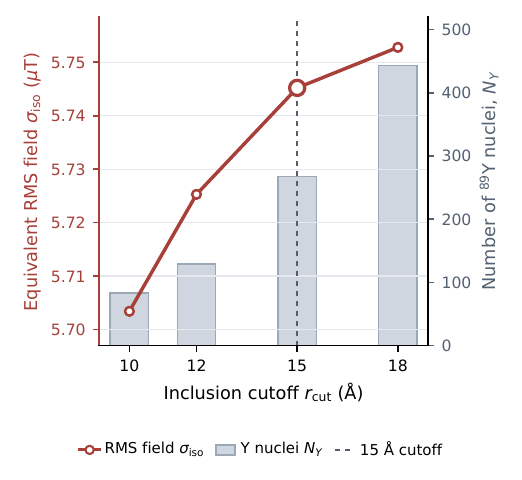}
\caption{$^{89}\mathrm{Y}$ magnetic-field covariance converging as the radial inclusion cutoff increases. The curve represents $\sigma_{\mathrm{iso}}=\sqrt{\mathrm{Tr}(\boldsymbol{\Sigma}_Y)/3}$, the equivalent isotropic RMS magnetic-field amplitude showed against the corresponding number of included Y nuclei. The vertical dashed line marks the point of convergence, the adopted 15~\AA cutoff, confining 267 nuclear spin with  $\sigma_{\mathrm{iso}}=5.745~\mu\mathrm{T}$.}
\label{fig:y-bath-convergence}
\end{figure}

An inclusion cutoff of $15.0$~\AA\ encompasses $N=267$ host $^{89}\mathrm{Y}$ spins and yields $\sigma_{\mathrm{iso}}=5.745~\mu\mathrm{T}$. Upon outward propagation of the dipolar summation radius to $18.0$~\AA\, 443 spins were incorporated but the RMS field amplitude grew only by $0.13\%$ which conveys strong convergence of the lattice-derived RMS field amplitude. Concurrently, in the same interval, the central correlation-time proxy changes by approximately $1.9\%$. We therefore adopt $15$~\AA\ as the foundational operational cutoff, with a characteristic $\tau_{\mathrm{RMS}}=12.444~\mathrm{ms}$.

\begin{figure*}[t]
\centering
\includegraphics[width=0.96\textwidth]{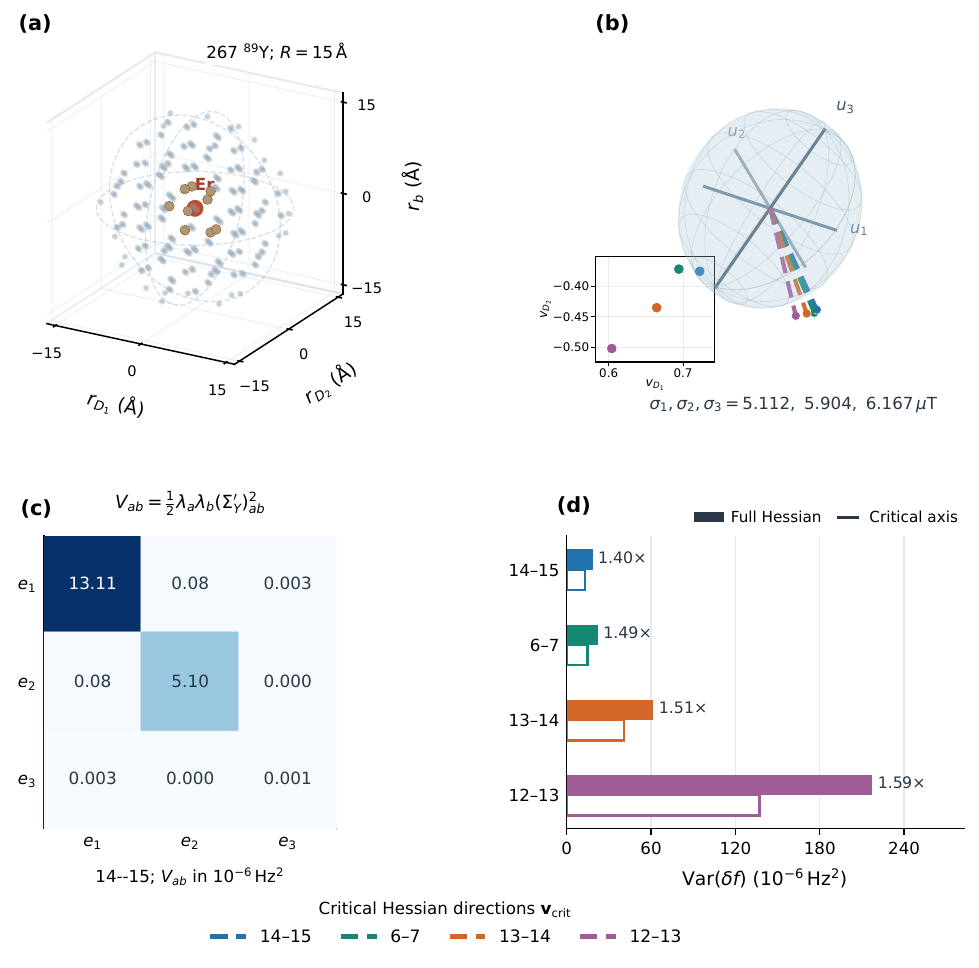}
\caption{Magnetic-field anisotropy and quadratic frequency-noise variance. (a) Spatial distribution of 267 surrounding $^{89}\mathrm{Y}$ nuclei within 15~\AA around central Site-2 Er ion. Highlighted are the nearby nuclei. (b)  Magnetic-field covariance ellipsoid. The semiaxes are proportional to the principal RMS fluctuations $\sigma_1=5.112$, $\sigma_2=5.904$, and $\sigma_3=6.167~\mu\mathrm{T}$. Critical Hessian directions of the four Pareto-optimal transitions are also shown. (c) Contributions to quadratic frequency-noise variance in the Hessian eigenbasis for the $\epsilon_{14}\leftrightarrow\epsilon_{15}$ transition. (d) Complete Hessian contraction compared with the critical-axis-only contribution for the four Pareto candidates.}
\label{fig:y-full-hessian}
\end{figure*}

Directional anisotropy strictly governs the evaluated magnetic-field covariance. In the optical $(D_1,D_2,b)$ frame, diagonalizing $\boldsymbol{\Sigma}_B$ yields principal field-noise RMS amplitudes of $5.112~\mu\mathrm{T}$, $5.904~\mu\mathrm{T}$, and $6.167~\mu\mathrm{T}$, with corresponding unit eigenvectors $\mathbf{u}_1=(-0.8860$, $-0.4636$, $+0.0071)$, $\mathbf{u}_2=(-0.2660$, $+0.4958$, $-0.8267)$, and $\mathbf{u}_3=(-0.3797$, $+0.7344$, $+0.5626)$. These unequal principal amplitudes show the anisotropy of the local nuclear magnetic field fluctuations and the eigenvectors identify that the principal directions are mutually orthogonal. Consequently, any single-axis approximation limited to $\mathbf{v}_{\mathrm{max}}$ proves to be fundamentally deficient because such a constrained projection truncates the complete quadratic frequency-noise variance by neglecting the remaining Hessian eigendirections and inherent cross-component field correlations.

Modeled Y-bath pure-dephasing times are evaluated from the full-Hessian second-cumulant framework, with the adopted exponential Y-field correlation function. For the five lowest-curvature transitions, the computed Kubo parameters decisively satisfy $\kappa\leq10^{-2}$. This anchors the dynamics deep within the fast-modulation regime. Their conditional, Y-exclusive magnetic coherence estimates are presented in Table~\ref{tab:y_only_coherence}.

\begin{table}[t]
\caption{Evaluated full-Hessian quadratic frequency-noise metrics and modeled Y-bath quadratic magnetic pure-dephasing times ($T_{\phi,Y}^{(2)}$) for the top five Site-2 ZEFOZ candidates. The Kubo parameter $\kappa$ confirms that all transitions reside deep within the fast-modulation limit.} \label{tab:y_only_coherence}
\begin{ruledtabular}
\begin{tabular}{l c c c c}
Transition & $S_{2,\mathrm{max}}$ & $\Delta_{\omega}/2\pi$ & $\kappa$ & $T_{\phi,Y}^{(2)}$ \\
 & (\text{Hz/mT}$^2$) & (\text{Hz}) & ($10^{-4}$) & (\text{s}) \\ 
\hline
$\epsilon_{14} \leftrightarrow \epsilon_{15}$ & 165.241 & $4.285 \times 10^{-3}$ & 1.675 & $2.217 \times 10^5$ \\ 
$\epsilon_{6} \leftrightarrow \epsilon_{7}$   & 173.068 & $4.698 \times 10^{-3}$ & 1.837 & $1.844 \times 10^5$ \\ 
$\epsilon_{13} \leftrightarrow \epsilon_{14}$ & 282.525 & $7.849 \times 10^{-3}$ & 3.068 & $6.608 \times 10^4$ \\ 
$\epsilon_{13} \leftrightarrow \epsilon_{15}$ & 432.162 & $1.159 \times 10^{-2}$ & 4.532 & $3.030 \times 10^4$ \\ 
$\epsilon_{12} \leftrightarrow \epsilon_{13}$ & 503.778 & $1.474 \times 10^{-2}$ & 5.764 & $1.873 \times 10^4$ \\ 
\end{tabular}
\end{ruledtabular}
\end{table}

The $\epsilon_{14}\leftrightarrow\epsilon_{15}$ transtion exhibits the longest modeled Y-bath quadratic pure-dephasing time that eclipses $2.2\times10^5~\mathrm{s}$. Across all five transition, the Kubo parameters sit resolutely on the order of $10^{-4}$. This indicates that the system operates deep within the fast-modulation regime under the adopted exponential correlation framework. These long pure-dephasing times reflect the profound suppression of quadratic magnetic dephasing at these specific identified ZEFOZ points. These projections stand solely as conditional Y-only estimates, excluding Er electronic-spin bath  and other relaxation or dephasing mechanisms.

\subsection{Lattice-Resolved Unpolarized Erbium Field Second Moment}
\label{subsec:er-sensitivity}

The lattice-resolved summation reveals a substantial occupation-averaged magnetic-field second moment associated with the dilute Er dopant ensemble. Extending the spatial inclusion cutoff to encompass over quarter-million possible substitutional sites saturates the equivalent per-component RMS field amplitude $\sigma_{\mathrm{iso,Er}}$ at geometric limit of $1.003~\mathrm{mT}$. The convergence of this unpolarized ensemble second moment at the concentration of 10 ppm is presented in Table~\ref{tab:er_lattice_convergence}.

\begin{table}[h]
\centering
\caption{Geometric convergence of the unpolarized $\mathrm{Er}$dipolar-field RMS scale for a $10~\mathrm{ppm}$ occupation probability as a function of the spatial inclusion cutoff.}
\label{tab:er_lattice_convergence}
\begin{tabular}{c c c c}
\hline\hline
Cutoff (\AA) & Possible Sites & Expected $\mathrm{Er}$ Count & $\sigma_{\mathrm{iso,Er}}$ ($\mu$T) \\
\hline
15.0  & 267    & 0.00267 & 999.793 \\
20.0  & 625    & 0.00625 & 1001.890 \\
30.0  & 2111   & 0.02111 & 1003.001 \\
50.0  & 9766   & 0.09766 & 1003.367 \\
100.0 & 78086  & 0.78086 & 1003.455 \\
150.0 & 263552 & 2.63552 & 1003.464 \\
\hline\hline
\end{tabular}
\end{table}
Scaling heavily, the resultant Er ensemble RMS field is approximately $175$ times larger than the $5.745~\mu\mathrm{T}$ Y nuclear-field scale. Yet, the distribution of this magnetic variance is highly skewed. Despite an expected dopant occupation of merely $0.00267$ Er ions within a 15~\AA radius, these immediate proximation ssites command roughly $99.27\%$ of the total Er field variance out to the 150~\AA cutoff. Such localization of the second moment reveals a statistical reality: rather than having a uniform mean-field background, the structural variance is disproportionately dominated by the rare, closely spaced dopant configurations. The corresponding covariance matrix in the CIF Cartesian frame is   
\begin{equation}
\boldsymbol{\Sigma}_{B,\mathrm{Er}} = \begin{bmatrix}
1.038 & -0.233 & 0.271 \\
-0.233 & 1.056 & 0.155 \\
0.271 & 0.155 & 0.927
\end{bmatrix} \mathrm{mT}^2.
\end{equation}
The emergence of nonzero off-diagonal tensor elements reflect directional cross-correlations arising from inter-axis coupling dictated by the crystallographic arrangement of substitutional sites and the anisotropic electronic $\mathbf{g}$ tensors. This is an inevitable consequence of the low local point-group symmetry.

This massive unpolarized Er field second moment highlights the potential importance of the dopant ensemble as a source of magnetic dephasing. But the physical reality of the environment intervenes. The intended operating regime of approximately 10 mK, together with the substantial magnetic bias fields required by ZEFOZ coordinates expects the Er electronic ensemble to strongly polarize, suppressing the Er-Er flip flop dynamics by reducing the population of oppositely oriented electronic spins \cite{bottger2006,budoyo2018,ulanowski2025}. Such polarization reduces the population of these exact spin configurations required for resonant exchange, effectively suppressing an important mechanism of Er-induced spectral diffusion. 

Therefore, the unpolarized second moment calculation establishes a hypothetical scale  for the ensemble-field variance scale and does not dictate the active dynamical noise expected at millikelvin temperatures. The static ensemble variance does not actually need to vanish. This provides a physical basis for neglecting Er-bath temporal fluctuations as a leading-order approximation in the Y-only coherence calculations.

\section{Microwave Coupling and Coherence--Control Trade-Off}
\label{sec:pareto}

\subsection{Screened Operating-Point Atlas}
\label{subsec:filtered-atlas}

Out of the 344 stationary points discovered in the mathematical root search, a mere 9 clear the five concurrent screening criteria. The requirements are: a minimum nominal bias field $\|\mathbf{B}_0\|\geq1.0~\mathrm{mT}$, a transition frequency $0<f\leq2000~\mathrm{MHz}$, a fast modulation parameter $\kappa\leq10^{-2}$ under the adopted Y-bath correlation model, a Modeled Y-bath quadratic magnetic pure-dephasing time $T_{\phi,Y}^{(2)}\geq1000~\mathrm{s}$, and a minimum optimized microwave coupling $M_{\mathrm{opt}}\geq10^{-4}~\mathrm{MHz/mT}$ \cite{rakonjac2020}.

These nine qualifying roots correspond to distinct transition pairs. To construct the final atlas, the procedure isolates a single representative per pair. This extraction prioritizes maximum Y-only coherence, followed by maximum microwave coupling and minimum bias field if competitive roots emerge. The physical coordinates and curvature metrics of these nine finalized ZEFOZ operating points are cataloged in Table~\ref{tab:atlas_physical}. Their corresponding coherence estimates and optimal microwave drive strengths follow in Table~\ref{tab:atlas_performance}.

\begin{table}[h]
\centering
\caption{Physical parameters, transition frequencies, and evaluated full-Hessian curvature metrics for the nine screened ZEFOZ operating points.}
\label{tab:atlas_physical}
\begin{tabular}{l c c c c}
\hline\hline
Transition & $\|\mathbf{B}_0\|$ (mT) & $f$ (MHz) & $S_{2,\mathrm{max}}$ (Hz/mT$^2$) & $\kappa/10^{-4}$ \\
\hline
$\epsilon_{14} \leftrightarrow \epsilon_{15}$ & 633.527 & 796.291 & 165.241 & 1.675 \\
$\epsilon_{6} \leftrightarrow \epsilon_{7}$   & 668.529 & 797.602 & 173.068 & 1.837 \\
$\epsilon_{13} \leftrightarrow \epsilon_{14}$ & 413.741 & 795.942 & 282.525 & 3.068 \\
$\epsilon_{13} \leftrightarrow \epsilon_{15}$ & 510.470 & 1592.182 & 432.162 & 4.532 \\
$\epsilon_{5} \leftrightarrow \epsilon_{7}$   & 513.260 & 1594.590 & 521.485 & 5.674 \\
$\epsilon_{12} \leftrightarrow \epsilon_{13}$ & 283.919 & 797.960 & 503.778 & 5.764 \\
$\epsilon_{5} \leftrightarrow \epsilon_{6}$   & 335.656 & 796.972 & 548.544 & 6.112 \\
$\epsilon_{12} \leftrightarrow \epsilon_{14}$ & 344.051 & 1593.872 & 743.672 & 8.292 \\
$\epsilon_{11} \leftrightarrow \epsilon_{13}$ & 233.187 & 1601.148 & 1448.957 & 16.890 \\
\hline\hline
\end{tabular}
\end{table}

\begin{table}[h]
\centering
\caption{Modeled Y-bath quadratic magnetic pure-dephasing time, optimized microwave coupling strength, and evaluated Pareto non-dominance status for the screened atlas candidates.}
\label{tab:atlas_performance}
\resizebox{\columnwidth}{!}{%
\begin{tabular}{l c c c c}
\hline\hline
Transition & $T_{\phi,Y}^{(2)}$ (s) & $M_{\mathrm{opt}}$ (MHz/mT) & $f_R/B_1$ (Hz/mT) & Pareto Optimal\\
\hline
$\epsilon_{14} \leftrightarrow \epsilon_{15}$ & $2.217 \times 10^5$ & $5.307 \times 10^{-2}$ & $5.307 \times 10^4$ & Yes \\
$\epsilon_{6} \leftrightarrow \epsilon_{7}$ & $1.844 \times 10^5$ & $5.638 \times 10^{-2}$ & $5.638 \times 10^4$ & Yes \\
$\epsilon_{13} \leftrightarrow \epsilon_{14}$ & $6.608 \times 10^4$ & $8.689 \times 10^{-2}$ & $8.689 \times 10^4$ & Yes \\
$\epsilon_{13} \leftrightarrow \epsilon_{15}$ & $3.030 \times 10^4$ & $5.687 \times 10^{-3}$ & $5.687 \times 10^3$ & No \\
$\epsilon_{5} \leftrightarrow \epsilon_{7}$ & $1.933 \times 10^4$ & $3.947 \times 10^{-3}$ & $3.947 \times 10^3$ & No \\
$\epsilon_{12} \leftrightarrow \epsilon_{13}$ & $1.873 \times 10^4$ & $1.248 \times 10^{-1}$ & $1.248 \times 10^5$ & Yes \\
$\epsilon_{5} \leftrightarrow \epsilon_{6}$ & $1.665 \times 10^4$ & $1.210 \times 10^{-1}$ & $1.210 \times 10^5$ & No \\
$\epsilon_{12} \leftrightarrow \epsilon_{14}$ & $9.050 \times 10^3$ & $1.061 \times 10^{-2}$ & $1.061 \times 10^4$ & No \\
$\epsilon_{11} \leftrightarrow \epsilon_{13}$ & $2.180 \times 10^3$ & $1.767 \times 10^{-2}$ & $1.767 \times 10^4$ & No \\
\hline\hline
\end{tabular}%
}
\end{table}

\subsection{Pareto Frontier and Candidate Comparison}
\label{subsec:coherence-control}

Upon screening the atlas, we encounter a trade-off between modeled Y-bath quadratic pure-dephasing time and optimized microwave transition strength. The $\epsilon_{14} \leftrightarrow \epsilon_{15}$ transition has the highest estimated modeled Y-bath quadratic pure-dephasing time among the screened candidates but on the other hand it yields a relatively limited microwave coupling metric of $5.307 \times 10^{-2}~\mathrm{MHz/mT}$. The  $\epsilon_{12} \leftrightarrow \epsilon_{13}$ transition has the highest microwave addressabilty in the atlas, reaching $M_{\mathrm{opt}} = 1.248 \times 10^{-1}~\mathrm{MHz/mT}$ which corresponds to an optimal Rabi frequency coefficient of $1.248 \times 10^5~\mathrm{Hz/mT}$ per unit drive field. 

We then evaluated non-dominance among the nine candidates to obtain a Pareto frontier having 4 transition pairs:  $\epsilon_{14} \leftrightarrow \epsilon_{15}$, $\epsilon_{6} \leftrightarrow \epsilon_{7}$, $\epsilon_{13} \leftrightarrow \epsilon_{14}$, and $\epsilon_{12} \leftrightarrow \epsilon_{13}$. Here, no objective can be improved without compromising the other, forming a non-dominated frontier in the coherence-microwave coupling plane. The other 5 transitions are dominated either by better coherence or microwave addressability, although adding few more constraints or on bias field, transition frequency, or experimental implementation might alter the candidate's practical relevance.

\begin{figure*}[t]
\centering
\includegraphics[width=0.92\textwidth]{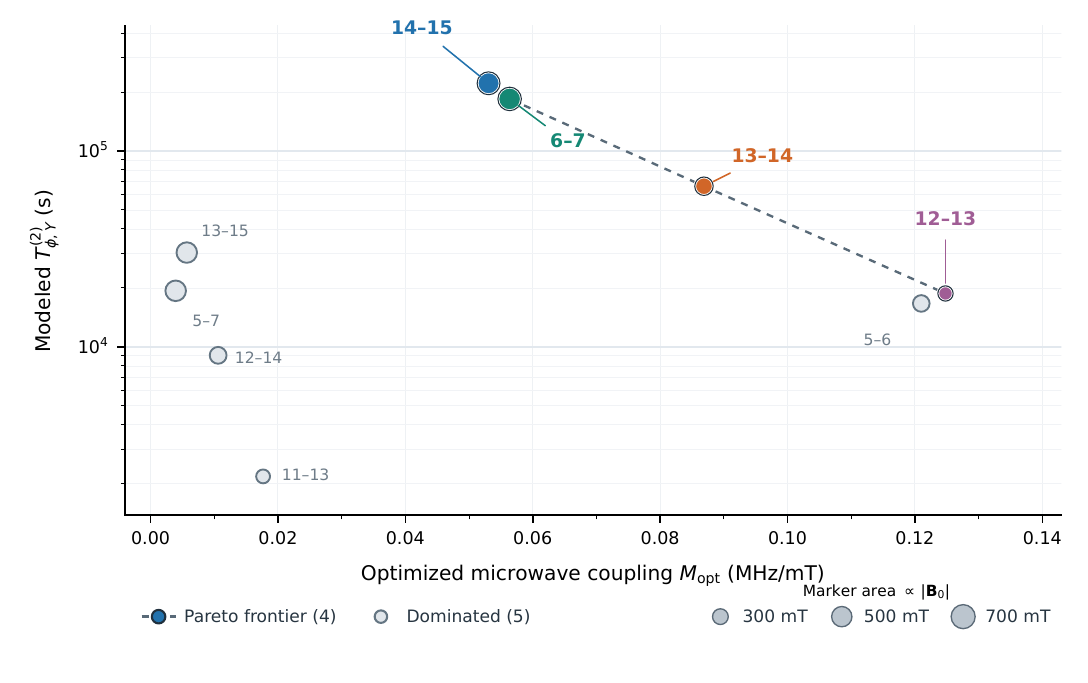}
\caption{Trade-off between the modeled Y-bath-induced quadratic magnetic pure-dephasing time $T_{\phi,Y}^{(2)}$ and the optimized microwave coupling $M_{\mathrm{opt}}$ for the nine screened ZEFOZ operating points. The four non-dominated transitions form the discrete Pareto frontier. In gray are the remaining five dominated candidates. Marker areas corresponds to its associated nominal magnetic bias-field magnitude $\|\mathbf{B}_0\|$. The dashed lines connect the Pareto candidates as a visual guide.}
\label{fig:pareto-frontier}
\end{figure*}

\section{Hamiltonian Uncertainty and Numerical Recalibration}
\label{sec:recalibration}

\subsection{Open-Loop Sensitivity and Numerical Recovery}
\label{subsec:recovery}

A physical memory will possess an effective spin Hamiltonian that slightly deviates from the nominal parameters \cite{welinski2017,galland2020} and the actual ZEFOZ coordinates will shift. To analyze operational robustness, we will subject the four Pareto-optimal Site-2 ZEFOZ candidates identified in Sec.~\ref{subsec:coherence-control} to an assumption-based Hamiltonian-parameter sensitivity analysis having 500 independently perturbed Hamiltonians per candidate. The tensors undergo perturbations with Gaussian relative standard deviations of $1\%$ for $\mathbf{A}$ and $\mathbf{g}$, and $2\%$ for $\mathbf{Q}$.

First we evaluate an open-loop operational scenario. In this, the magnetic field bias is left unchanged from its nominal coordinate, subject to an assumed zero-mean Gaussian field-setting error with a standard deviation of $0.10~\mathrm{mT}$ per spatial axis. We generally observe that the critical ZEFOZ condition is destroyed under the combination of this deviated Hamiltonian and placement error, which results in an uncompensated first order sensitivity gradient  ($\mathbf{S}_1 \neq \mathbf{0}$) reintroducing the rapid linear dephasing. By implementing the fast-modulation linear-dephasing rate approximation, the median inverse-rate timescales lie in the range of 19.8 to 71.8 ms among the four candidates. None of the 500 open-loop samples per candidate have a rate-based magnetic-dephasing estimate above 1000 s. These timescales are not exact coherence lifetimes, they just quantify the increase in the first-order magnetic sensitivity. The residual linear frequency noise is not independently established to satisfy the fast-modulation condition. 

However we can locate the shifted operating coordinates numerically by recalibrating the perturbed Hamiltonian. The step-limited regularized Newton solver converges to a ZEFOZ root  ($\|\mathbf{S}_1\|<10^{-8}~\mathrm{MHz/mT}$) for all the sampled Hamiltonians per candidate and the recovered solution also satifies the quadratic-noise fast-modulation criterion  $\kappa<10^{-2}$. Keeping the Y-bath covariance and correlation-time proxy fixed, and assuming the field is placed ideally post recalibration, the median modeled Y-bath quadratic pure-dephasing times regain the values ranging from $1.866\times10^4$ to $2.233\times10^5~\mathrm{s}$.

\subsection{Field-Relocation Requirements}
\label{subsec:field-relocation}

It must be acknowledged that even though algorithmic convergence does indicate mathematical recovery, a practical experiment apparatus has to be capable of physically traversing the required vector relocation  ($\Delta B$). Table~\ref{tab:recalibration_shift} summarizes the statistical distribution of the required field shifts.

\begin{table}[h]
\centering
\caption{Vector magnetic-field relocation magnitude required to recover the ZEFOZ condition of the perturbed Hamiltonian.}
\label{tab:recalibration_shift}
\begin{tabular}{l c c}
\hline\hline
Transition & Median $\Delta B$ (mT) & $84.1^{\mathrm{st}}$ Percentile $\Delta B$ (mT) \\
\hline
$\epsilon_{14} \leftrightarrow \epsilon_{15}$ & 14.099 & 27.025 \\
$\epsilon_{6} \leftrightarrow \epsilon_{7}$   & 15.097 & 27.965 \\
$\epsilon_{13} \leftrightarrow \epsilon_{14}$ & 8.702  & 15.486 \\
$\epsilon_{12} \leftrightarrow \epsilon_{13}$ & 6.036  & 10.950 \\
\hline\hline
\end{tabular}
\end{table}

The $\epsilon_{14} \leftrightarrow \epsilon_{15}$ transition exhibiting longest modeled pure-dephasing time needs a median field relocation of roughly $14.1~\mathrm{mT}$, with  $84.1\%$ perturbed samples requiring a shift of less than $27.1~\mathrm{mT}$. The better microwave-addressable $\epsilon_{12} \leftrightarrow \epsilon_{13}$ transition demands a smaller shift, having the $84.1^{\mathrm{st}}$ percentile boundary near $11.0~\mathrm{mT}$. This demonstrates the numerical trackability of these transitions under the implemented Hamiltonian perturbations. Experimental recovery also depends on available vector-field calibration range and accuracy.
\begin{figure*}[t]
\centering
\includegraphics[width=0.96\textwidth]{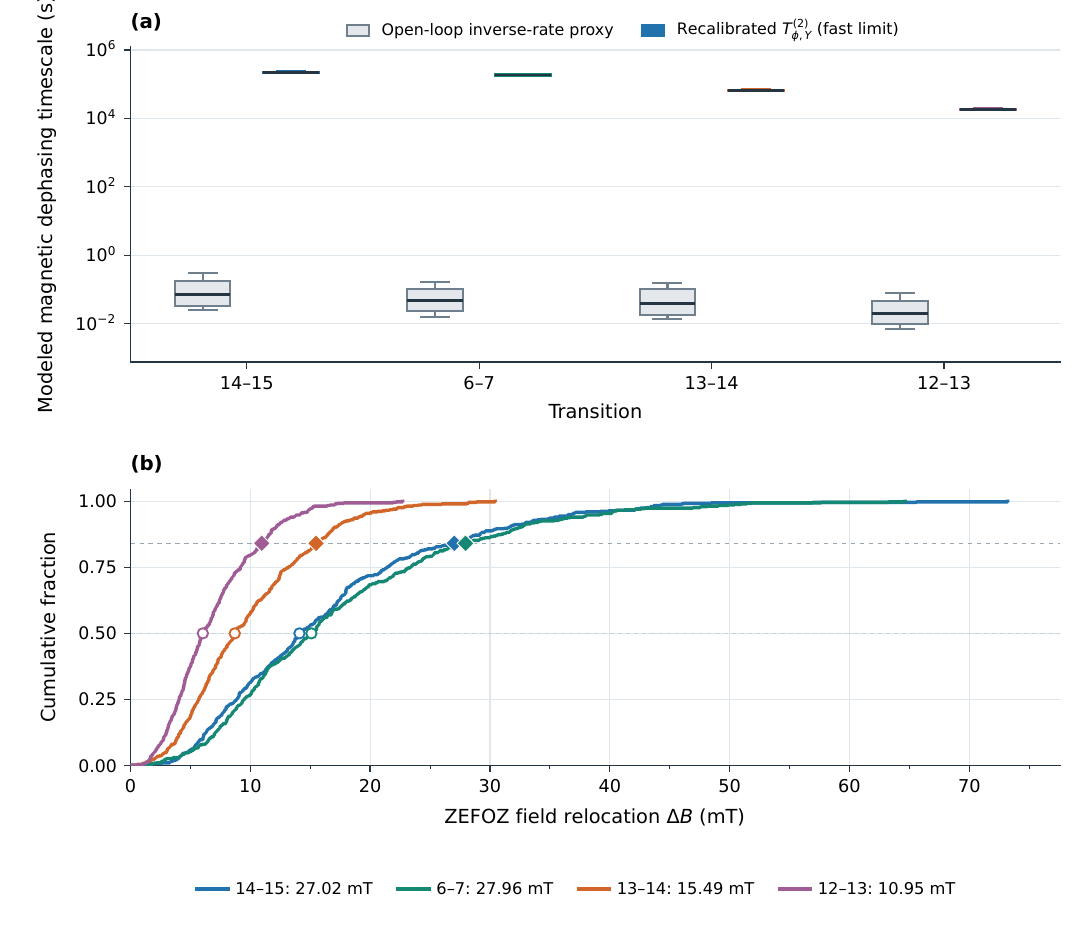}
\caption{Open-loop magnetic sensitivity and the numerical recovery of ZEFOZ points for the sampled Hamiltonian-parameter uncertainties. (a) Open-loop inverse-rate magnetic-dephasing estimates are compared with their corresponding recalibrated Y-bath quadratic pure-dephasing times for the respective four Pareto-optimal transition. Boxes show interquartile ranges, horizontal lines mark medians, and whiskers represent the 15.9th and 84.1st percentiles. (b) Cumulative empirical distributions of the vector magnetic-field relocation necessary for numerical recalibration. Hollow circles indicate median relocations, while filled diamonds mark the 84.1st percentiles. 500 independent perturbed Hamiltonians are used to evaluate each transition.}
\label{fig:recalibration}
\end{figure*}
\section{Discussion}
\label{sec:discussion}

\subsection{Full-Hessian Treatment in Relation to Prior ZEFOZ Calculations}
\label{subsec:relation-prior}

Identification of ZEFOZ transitions by minimizing the first-order Zeeman sensitivity and assessing the residual quadratic response has already been established in previous literature\cite{mcauslan2012}. Matsuura et. al.\cite{g23x-14ks} located the Site 2 operating point near $633.52~\mathrm{mT}$ and its coherence using Monte-Carlo derived magnetic-field fluctuations and a quadratic form involving the second-order Zeeman sensitivity tensor. The distinction lies not in the introduction of a tensorial curvature description itself; the current treatment presents the direct multiplication of the full frequency Hessian with the lattice-derived magnetic-field covariance tensor, to then perform a temporal-correlation-based coherence calculation. 

The off-diagonal curvature terms and the relative orientations of the frequency Hessian and anisotropic $^{89}\mathrm{Y}$ magnetic-field covariance tensor is taken into account and kept intact in this full-Hessian framework. When we observe off-diagonal covariance elements, it is crucial to describe these correlations between Cartesian components of the local magnetic field caused by lattice geometry rather than correlations between individual statistically independent nuclear spins. Contracting these tensors yields the quadratic frequency-noise variance, and we need not impose an isotropic field distribution or reduce the curvature to a single principal direction. The resulting fast-modulation parameters ($\kappa$) and Y-bath pure-dephasing estimates ($T_{\phi,Y}^{(2)}$) relies on the adopted Gaussian field statistics,exponential temporal-correlation model, and second-cumulant approximation.

\subsection{Model Limitations and Experimental Validation}
\label{subsec:model-boundaries}

We can only justify such exceptional modeled Y-bath magnetic pure-dephasing times upon the consideration of the dependence on the adopted magnetic-noise and temporal-correlation models. The motional-narrowing analysis is conditional on the adopted temporal-correlation model and most of our estimates pertaining to the coherence lifetimes relies on the same. For the Er-electronic spin ensemble whose unpolarized lattice-resolved second moment lays out an RMS field scale of roughly $1.003~\mathrm{mT}$, we must consider the intended operating temperature of approximately $10~\mathrm{mK}$, and expect significant DC magnetic bias fields to strongly polarize the Er electronic spins, suppressing resonant Er-Er flip-flops and the associated spectral diffusion \cite{bottger2006,budoyo2018,ulanowski2025}. This serves as grounds to disregard the dynamical Er-bath dephasing as a leading-order low-temperature approximation. The unpolarized ensemble second moment characterizes the structural capacity for magnetic-field variance but does not quantify the residual dynamical noise under polarized operating conditions.

More importantly, the simulation model used in this study only tracks pure magnetic dephasing time and does not explicitly calculate spin-lattice relaxation ($T_1$).
The experimentally observable coherence time may be shorter than the modeled Y-bath pure-dephasing time because population relaxation and other dephasing mechanisms can contribute independently. The targeted millikelvin regime suppresses thermally driven spin relaxation, but population relaxation will independently limit coherence through $T_2\leq 2T_1$ \cite{abragam1961,budoyo2018}. 

Finally, for the recalibration model, even though the perturbation analysis demonstrates numerical recovery of shifted ZEFOZ coordinates, each sampled Hamiltonian represents a single set of uniformly perturbed parameters rather than a spatial distribution of spatial environments. In a macroscopic crystal, physical stress/inhomogeneous strain can warp the effective spin Hamiltonian \cite{welinski2017,galland2020}, such that a uniform external magnetic field fails to satisfy the ZEFOZ condition throughout the entire ensemble. As a result, the calculated field-relocation distribution establish numerical trackability for the subjected sampled uncertainties but the vector-field precision or crystal homogeneity required for ensemble-based spin-wave storage cannot be independently determined. 

\section{Conclusion}
\label{sec:conclusion}

In this work we have evaluated an operating landscape for Site-2 $^{167}\mathrm{Er}^{3+}:\mathrm{Y}_2\mathrm{SiO}_5$ quantum memories. We analyze the full-Hessian spin-Hamiltonian and further interpret it with the anisotropic magnetic-field covariance of the native $^{89}\mathrm{Y}$ lattice and an explicit temporal-correlation model. To show a trade-off between Y-only pure magnetic-dephasing time and microwave control authority, we isolate a Pareto-optimal frontier of transition pairs which balances these both parameters. The $\epsilon_{14} \leftrightarrow \epsilon_{15}$ transition maintains a long modeled Y-bath quadratic pure-dephasing time of  $2.2 \times 10^5~\mathrm{s}$ and the $\epsilon_{12} \leftrightarrow \epsilon_{13}$ coordinates exhibits superior hardware addressability with a Rabi frequency coefficient of  $1.248 \times 10^5~\mathrm{Hz/mT}$ associated with it. We demonstrate by Hamiltonian-uncertainty analysis, the possibility of numerical recovery of the four Pareto-optimal transitions for every sampled parameter perturbation with the $84.1^{\mathrm{st}}$-percentile field relocations that ranges from $10.95$ to $27.97~\mathrm{mT}$. With these findings we present physically screened operating-points for experimental vector-field searches, with Y-only pure magnetic-dephasing lifetime estimates conditional on the adopted nuclear-bath dynamics and the suppression of Er-induced spectral diffusion at the intended millikelvin operating temperatures.

\begin{acknowledgments}
I would like express my sincere gratitude to Professor Anu Venugopalan, Professor Harish Parthasarathy, Professor Anirban Pathak, and Professor Sandeep Mishra for their support across various stages of this paper. I also acknowledge the JAX development team at Google DeepMind for the open-source JAX framework. Automatic-differentiation and numerical-computation capabilities were crucial for evaluating the spin-Hamiltonian gradients and Hessians used in this work.
\end{acknowledgments}

\section*{Data and Code Availability}
The numerical results that support this study are presented in the manuscript. The calculations were performed using custom Python code and Jupyter notebooks.  The data and code are available from the author upon reasonable request.

\appendix

\section{Spin Hamiltonian Parameters}
\label{app:hamiltonian}

The Site-2 spin-Hamiltonian tensors adopted from Ref.~\cite{wang2023}, expressed in the $(D_1,D_2,b)$ frame, are
\begin{equation}
\mathbf{A}_{\mathrm{site2}}=
\begin{pmatrix}
-1570 & 224 & -131\\
224 & -17 & -15\\
-131 & -15 & 141
\end{pmatrix}
\ \mathrm{MHz},
\end{equation}
\begin{equation}
\mathbf{Q}_{\mathrm{site2}}=
\begin{pmatrix}
-9.8 & -21 & -0.4\\
-21 & -16 & -12.4\\
-0.4 & -12.4 & 25.8
\end{pmatrix}
\ \mathrm{MHz},
\end{equation}
\begin{equation}
\mathbf{g}_{\mathrm{site2}}=
\begin{pmatrix}
14.44 & -1.76 & 2.35\\
-1.76 & 1.91 & -0.46\\
2.35 & -0.46 & 1.424
\end{pmatrix}.
\end{equation}
The numerical calculations use the symmetry-related subsite tensors $\mathbf{T}_0=\mathbf{R}_{C_2}\mathbf{T}_{\mathrm{site2}}\mathbf{R}_{C_2}^{\mathsf T}$, where $\mathbf{R}_{C_2}=\operatorname{diag}(-1,-1,1)$ and $\mathbf{T}\in\{\mathbf{A},\mathbf{Q},\mathbf{g}\}$.

\bibliography{references}

\begin{thebibliography}{26}%
\makeatletter
\providecommand \@ifxundefined [1]{%
 \@ifx{#1\undefined}
}%
\providecommand \@ifnum [1]{%
 \ifnum #1\expandafter \@firstoftwo
 \else \expandafter \@secondoftwo
 \fi
}%
\providecommand \@ifx [1]{%
 \ifx #1\expandafter \@firstoftwo
 \else \expandafter \@secondoftwo
 \fi
}%
\providecommand \natexlab [1]{#1}%
\providecommand \enquote  [1]{``#1''}%
\providecommand \bibnamefont  [1]{#1}%
\providecommand \bibfnamefont [1]{#1}%
\providecommand \citenamefont [1]{#1}%
\providecommand \href@noop [0]{\@secondoftwo}%
\providecommand \href [0]{\begingroup \@sanitize@url \@href}%
\providecommand \@href[1]{\@@startlink{#1}\@@href}%
\providecommand \@@href[1]{\endgroup#1\@@endlink}%
\providecommand \@sanitize@url [0]{\catcode `\\12\catcode `\$12\catcode `\&12\catcode `\#12\catcode `\^12\catcode `\_12\catcode `\%12\relax}%
\providecommand \@@startlink[1]{}%
\providecommand \@@endlink[0]{}%
\providecommand \url  [0]{\begingroup\@sanitize@url \@url }%
\providecommand \@url [1]{\endgroup\@href {#1}{\urlprefix }}%
\providecommand \urlprefix  [0]{URL }%
\providecommand \Eprint [0]{\href }%
\providecommand \doibase [0]{https://doi.org/}%
\providecommand \selectlanguage [0]{\@gobble}%
\providecommand \bibinfo  [0]{\@secondoftwo}%
\providecommand \bibfield  [0]{\@secondoftwo}%
\providecommand \translation [1]{[#1]}%
\providecommand \BibitemOpen [0]{}%
\providecommand \bibitemStop [0]{}%
\providecommand \bibitemNoStop [0]{.\EOS\space}%
\providecommand \EOS [0]{\spacefactor3000\relax}%
\providecommand \BibitemShut  [1]{\csname bibitem#1\endcsname}%
\let\auto@bib@innerbib\@empty
\bibitem [{\citenamefont {Kimble}(2008)}]{kimble2008}%
  \BibitemOpen
  \bibfield  {author} {\bibinfo {author} {\bibfnamefont {H.~J.}\ \bibnamefont {Kimble}},\ }\bibfield  {title} {\bibinfo {title} {The quantum internet},\ }\href {https://doi.org/10.1038/nature07127} {\bibfield  {journal} {\bibinfo  {journal} {Nature}\ }\textbf {\bibinfo {volume} {453}},\ \bibinfo {pages} {1023} (\bibinfo {year} {2008})}\BibitemShut {NoStop}%
\bibitem [{\citenamefont {Afzelius}\ \emph {et~al.}(2010)\citenamefont {Afzelius}, \citenamefont {Usmani}, \citenamefont {Amari}, \citenamefont {Lauritzen}, \citenamefont {Walther}, \citenamefont {Simon}, \citenamefont {Sangouard}, \citenamefont {Min{\'a}{\v r}}, \citenamefont {de~Riedmatten}, \citenamefont {Gisin},\ and\ \citenamefont {Kr{"o}ll}}]{afzelius2010}%
  \BibitemOpen
  \bibfield  {author} {\bibinfo {author} {\bibfnamefont {M.}~\bibnamefont {Afzelius}}, \bibinfo {author} {\bibfnamefont {I.}~\bibnamefont {Usmani}}, \bibinfo {author} {\bibfnamefont {A.}~\bibnamefont {Amari}}, \bibinfo {author} {\bibfnamefont {B.}~\bibnamefont {Lauritzen}}, \bibinfo {author} {\bibfnamefont {A.}~\bibnamefont {Walther}}, \bibinfo {author} {\bibfnamefont {C.}~\bibnamefont {Simon}}, \bibinfo {author} {\bibfnamefont {N.}~\bibnamefont {Sangouard}}, \bibinfo {author} {\bibfnamefont {J.}~\bibnamefont {Min{\'a}{\v r}}}, \bibinfo {author} {\bibfnamefont {H.}~\bibnamefont {de~Riedmatten}}, \bibinfo {author} {\bibfnamefont {N.}~\bibnamefont {Gisin}},\ and\ \bibinfo {author} {\bibfnamefont {S.}~\bibnamefont {Kr{"o}ll}},\ }\bibfield  {title} {\bibinfo {title} {Demonstration of atomic frequency comb memory for light with spin-wave storage},\ }\href {https://doi.org/10.1103/PhysRevLett.104.040503} {\bibfield  {journal} {\bibinfo  {journal} {Phys. Rev. Lett.}\ }\textbf {\bibinfo {volume} {104}},\ \bibinfo
  {pages} {040503} (\bibinfo {year} {2010})}\BibitemShut {NoStop}%
\bibitem [{\citenamefont {Jiang}\ \emph {et~al.}(2023)\citenamefont {Jiang}, \citenamefont {Xue}, \citenamefont {He}, \citenamefont {An}, \citenamefont {Zheng}, \citenamefont {Xu}, \citenamefont {Xie}, \citenamefont {Lu}, \citenamefont {Zhu},\ and\ \citenamefont {Ma}}]{jiang2023}%
  \BibitemOpen
  \bibfield  {author} {\bibinfo {author} {\bibfnamefont {M.-H.}\ \bibnamefont {Jiang}}, \bibinfo {author} {\bibfnamefont {W.}~\bibnamefont {Xue}}, \bibinfo {author} {\bibfnamefont {Q.}~\bibnamefont {He}}, \bibinfo {author} {\bibfnamefont {Y.-Y.}\ \bibnamefont {An}}, \bibinfo {author} {\bibfnamefont {X.}~\bibnamefont {Zheng}}, \bibinfo {author} {\bibfnamefont {W.-J.}\ \bibnamefont {Xu}}, \bibinfo {author} {\bibfnamefont {Y.-B.}\ \bibnamefont {Xie}}, \bibinfo {author} {\bibfnamefont {Y.}~\bibnamefont {Lu}}, \bibinfo {author} {\bibfnamefont {S.}~\bibnamefont {Zhu}},\ and\ \bibinfo {author} {\bibfnamefont {X.-S.}\ \bibnamefont {Ma}},\ }\bibfield  {title} {\bibinfo {title} {Quantum storage of entangled photons at telecom wavelengths in a crystal},\ }\href {https://doi.org/10.1038/s41467-023-42741-1} {\bibfield  {journal} {\bibinfo  {journal} {Nat. Commun.}\ }\textbf {\bibinfo {volume} {14}},\ \bibinfo {pages} {6995} (\bibinfo {year} {2023})}\BibitemShut {NoStop}%
\bibitem [{\citenamefont {Ran{\v c}i{\'c}}\ \emph {et~al.}(2018)\citenamefont {Ran{\v c}i{\'c}}, \citenamefont {Hedges}, \citenamefont {Ahlefeldt},\ and\ \citenamefont {Sellars}}]{rancic2018}%
  \BibitemOpen
  \bibfield  {author} {\bibinfo {author} {\bibfnamefont {M.}~\bibnamefont {Ran{\v c}i{\'c}}}, \bibinfo {author} {\bibfnamefont {M.~P.}\ \bibnamefont {Hedges}}, \bibinfo {author} {\bibfnamefont {R.~L.}\ \bibnamefont {Ahlefeldt}},\ and\ \bibinfo {author} {\bibfnamefont {M.~J.}\ \bibnamefont {Sellars}},\ }\bibfield  {title} {\bibinfo {title} {Coherence time of over a second in a telecom-compatible quantum memory storage material},\ }\href {https://doi.org/10.1038/nphys4254} {\bibfield  {journal} {\bibinfo  {journal} {Nat. Phys.}\ }\textbf {\bibinfo {volume} {14}},\ \bibinfo {pages} {50} (\bibinfo {year} {2018})}\BibitemShut {NoStop}%
\bibitem [{\citenamefont {Wang}\ \emph {et~al.}(2023)\citenamefont {Wang}, \citenamefont {Chen}, \citenamefont {Longdell},\ and\ \citenamefont {Zhang}}]{wang2023}%
  \BibitemOpen
  \bibfield  {author} {\bibinfo {author} {\bibfnamefont {S.-J.}\ \bibnamefont {Wang}}, \bibinfo {author} {\bibfnamefont {Y.-H.}\ \bibnamefont {Chen}}, \bibinfo {author} {\bibfnamefont {J.~J.}\ \bibnamefont {Longdell}},\ and\ \bibinfo {author} {\bibfnamefont {X.}~\bibnamefont {Zhang}},\ }\bibfield  {title} {\bibinfo {title} {Hyperfine states of erbium doped yttrium orthosilicate for long-coherence-time quantum memories},\ }\href {https://doi.org/10.1016/j.jlumin.2023.119935} {\bibfield  {journal} {\bibinfo  {journal} {J. Lumin.}\ }\textbf {\bibinfo {volume} {262}},\ \bibinfo {pages} {119935} (\bibinfo {year} {2023})}\BibitemShut {NoStop}%
\bibitem [{\citenamefont {Matsuura}\ \emph {et~al.}(2026)\citenamefont {Matsuura}, \citenamefont {Yasui}, \citenamefont {Kaji}, \citenamefont {Sasakura}, \citenamefont {Tawara},\ and\ \citenamefont {Adachi}}]{g23x-14ks}%
  \BibitemOpen
  \bibfield  {author} {\bibinfo {author} {\bibfnamefont {K.}~\bibnamefont {Matsuura}}, \bibinfo {author} {\bibfnamefont {S.}~\bibnamefont {Yasui}}, \bibinfo {author} {\bibfnamefont {R.}~\bibnamefont {Kaji}}, \bibinfo {author} {\bibfnamefont {H.}~\bibnamefont {Sasakura}}, \bibinfo {author} {\bibfnamefont {T.}~\bibnamefont {Tawara}},\ and\ \bibinfo {author} {\bibfnamefont {S.}~\bibnamefont {Adachi}},\ }\bibfield  {title} {\bibinfo {title} {Exploration of optimal hyperfine transitions for spin-wave storage in {$^{167}$Er$^{3+}$:Y$_2$SiO$_5$}},\ }\href {https://doi.org/10.1103/g23x-14ks} {\bibfield  {journal} {\bibinfo  {journal} {Phys. Rev. B}\ }\textbf {\bibinfo {volume} {113}},\ \bibinfo {pages} {085421} (\bibinfo {year} {2026})}\BibitemShut {NoStop}%
\bibitem [{\citenamefont {B{"o}ttger}\ \emph {et~al.}(2006)\citenamefont {B{"o}ttger}, \citenamefont {Thiel}, \citenamefont {Sun},\ and\ \citenamefont {Cone}}]{bottger2006}%
  \BibitemOpen
  \bibfield  {author} {\bibinfo {author} {\bibfnamefont {T.}~\bibnamefont {B{"o}ttger}}, \bibinfo {author} {\bibfnamefont {C.~W.}\ \bibnamefont {Thiel}}, \bibinfo {author} {\bibfnamefont {Y.}~\bibnamefont {Sun}},\ and\ \bibinfo {author} {\bibfnamefont {R.~L.}\ \bibnamefont {Cone}},\ }\bibfield  {title} {\bibinfo {title} {Optical decoherence and spectral diffusion at 1.5 {$\mu$m} in {Er$^{3+}$:Y$_2$SiO$_5$} versus magnetic field, temperature, and {Er$^{3+}$} concentration},\ }\href {https://doi.org/10.1103/PhysRevB.73.075101} {\bibfield  {journal} {\bibinfo  {journal} {Phys. Rev. B}\ }\textbf {\bibinfo {volume} {73}},\ \bibinfo {pages} {075101} (\bibinfo {year} {2006})}\BibitemShut {NoStop}%
\bibitem [{\citenamefont {Rakonjac}\ \emph {et~al.}(2020)\citenamefont {Rakonjac}, \citenamefont {Chen}, \citenamefont {Horvath},\ and\ \citenamefont {Longdell}}]{rakonjac2020}%
  \BibitemOpen
  \bibfield  {author} {\bibinfo {author} {\bibfnamefont {J.~V.}\ \bibnamefont {Rakonjac}}, \bibinfo {author} {\bibfnamefont {Y.-H.}\ \bibnamefont {Chen}}, \bibinfo {author} {\bibfnamefont {S.~P.}\ \bibnamefont {Horvath}},\ and\ \bibinfo {author} {\bibfnamefont {J.~J.}\ \bibnamefont {Longdell}},\ }\bibfield  {title} {\bibinfo {title} {Long spin coherence times in the ground state and in an optically excited state of {$^{167}$Er$^{3+}$:Y$_2$SiO$_5$} at zero magnetic field},\ }\href {https://doi.org/10.1103/PhysRevB.101.184430} {\bibfield  {journal} {\bibinfo  {journal} {Phys. Rev. B}\ }\textbf {\bibinfo {volume} {101}},\ \bibinfo {pages} {184430} (\bibinfo {year} {2020})}\BibitemShut {NoStop}%
\bibitem [{\citenamefont {Ulanowski}\ \emph {et~al.}(2025)\citenamefont {Ulanowski}, \citenamefont {Kuijpers}, \citenamefont {Merkel}, \citenamefont {Holz{"a}pfel},\ and\ \citenamefont {Reiserer}}]{ulanowski2025}%
  \BibitemOpen
  \bibfield  {author} {\bibinfo {author} {\bibfnamefont {A.}~\bibnamefont {Ulanowski}}, \bibinfo {author} {\bibfnamefont {O.}~\bibnamefont {Kuijpers}}, \bibinfo {author} {\bibfnamefont {B.}~\bibnamefont {Merkel}}, \bibinfo {author} {\bibfnamefont {A.}~\bibnamefont {Holz{"a}pfel}},\ and\ \bibinfo {author} {\bibfnamefont {A.}~\bibnamefont {Reiserer}},\ }\bibfield  {title} {\bibinfo {title} {Cavity-enhanced spectroscopy of individual nuclear spins in a dense bath},\ }\href {https://doi.org/10.1103/PRXQuantum.6.020344} {\bibfield  {journal} {\bibinfo  {journal} {PRX Quantum}\ }\textbf {\bibinfo {volume} {6}},\ \bibinfo {pages} {020344} (\bibinfo {year} {2025})}\BibitemShut {NoStop}%
\bibitem [{\citenamefont {Fraval}\ \emph {et~al.}(2004)\citenamefont {Fraval}, \citenamefont {Sellars},\ and\ \citenamefont {Longdell}}]{fraval2004}%
  \BibitemOpen
  \bibfield  {author} {\bibinfo {author} {\bibfnamefont {E.}~\bibnamefont {Fraval}}, \bibinfo {author} {\bibfnamefont {M.~J.}\ \bibnamefont {Sellars}},\ and\ \bibinfo {author} {\bibfnamefont {J.~J.}\ \bibnamefont {Longdell}},\ }\bibfield  {title} {\bibinfo {title} {Method of extending hyperfine coherence times in {Pr$^{3+}$:Y$_2$SiO$_5$}},\ }\href {https://doi.org/10.1103/PhysRevLett.92.077601} {\bibfield  {journal} {\bibinfo  {journal} {Phys. Rev. Lett.}\ }\textbf {\bibinfo {volume} {92}},\ \bibinfo {pages} {077601} (\bibinfo {year} {2004})}\BibitemShut {NoStop}%
\bibitem [{\citenamefont {McAuslan}\ \emph {et~al.}(2012)\citenamefont {McAuslan}, \citenamefont {Bartholomew}, \citenamefont {Sellars},\ and\ \citenamefont {Longdell}}]{mcauslan2012}%
  \BibitemOpen
  \bibfield  {author} {\bibinfo {author} {\bibfnamefont {D.~L.}\ \bibnamefont {McAuslan}}, \bibinfo {author} {\bibfnamefont {J.~G.}\ \bibnamefont {Bartholomew}}, \bibinfo {author} {\bibfnamefont {M.~J.}\ \bibnamefont {Sellars}},\ and\ \bibinfo {author} {\bibfnamefont {J.~J.}\ \bibnamefont {Longdell}},\ }\bibfield  {title} {\bibinfo {title} {Reducing decoherence in optical and spin transitions in rare-earth-metal-ion-doped materials},\ }\href {https://doi.org/10.1103/PhysRevA.85.032339} {\bibfield  {journal} {\bibinfo  {journal} {Phys. Rev. A}\ }\textbf {\bibinfo {volume} {85}},\ \bibinfo {pages} {032339} (\bibinfo {year} {2012})}\BibitemShut {NoStop}%
\bibitem [{\citenamefont {Kubo}(1962)}]{kubo1962}%
  \BibitemOpen
  \bibfield  {author} {\bibinfo {author} {\bibfnamefont {R.}~\bibnamefont {Kubo}},\ }\bibfield  {title} {\bibinfo {title} {Generalized cumulant expansion method},\ }\href {https://doi.org/10.1143/JPSJ.17.1100} {\bibfield  {journal} {\bibinfo  {journal} {J. Phys. Soc. Jpn.}\ }\textbf {\bibinfo {volume} {17}},\ \bibinfo {pages} {1100} (\bibinfo {year} {1962})}\BibitemShut {NoStop}%
\bibitem [{\citenamefont {{National Institute of Standards and Technology}}({\natexlab{a}})}]{nistEr}%
  \BibitemOpen
  \bibfield  {author} {\bibinfo {author} {\bibnamefont {{National Institute of Standards and Technology}}},\ }\href {https://physics.nist.gov/PhysRefData/Handbook/Tables/erbiumtable1.htm} {\bibinfo {title} {Atomic data for erbium ({Er})}},\ \bibinfo {howpublished} {NIST Atomic Spectra Database Handbook} ({\natexlab{a}})\BibitemShut {NoStop}%
\bibitem [{\citenamefont {Sun}\ \emph {et~al.}(2008)\citenamefont {Sun}, \citenamefont {B{"o}ttger}, \citenamefont {Thiel},\ and\ \citenamefont {Cone}}]{sun2008}%
  \BibitemOpen
  \bibfield  {author} {\bibinfo {author} {\bibfnamefont {Y.}~\bibnamefont {Sun}}, \bibinfo {author} {\bibfnamefont {T.}~\bibnamefont {B{"o}ttger}}, \bibinfo {author} {\bibfnamefont {C.~W.}\ \bibnamefont {Thiel}},\ and\ \bibinfo {author} {\bibfnamefont {R.~L.}\ \bibnamefont {Cone}},\ }\bibfield  {title} {\bibinfo {title} {Magnetic {$g$} tensors for the {$^{4}I_{15/2}$} and {$^{4}I_{13/2}$} states of {Er$^{3+}$:Y$_2$SiO$_5$}},\ }\href {https://doi.org/10.1103/PhysRevB.77.085124} {\bibfield  {journal} {\bibinfo  {journal} {Phys. Rev. B}\ }\textbf {\bibinfo {volume} {77}},\ \bibinfo {pages} {085124} (\bibinfo {year} {2008})}\BibitemShut {NoStop}%
\bibitem [{\citenamefont {Bradbury}\ \emph {et~al.}(2018)\citenamefont {Bradbury}, \citenamefont {Frostig}, \citenamefont {Hawkins}, \citenamefont {Johnson}, \citenamefont {Katariya}, \citenamefont {Leary}, \citenamefont {Maclaurin}, \citenamefont {Necula}, \citenamefont {Paszke}, \citenamefont {Vander{P}las}, \citenamefont {Wanderman-{M}ilne},\ and\ \citenamefont {Zhang}}]{jax2018github}%
  \BibitemOpen
  \bibfield  {author} {\bibinfo {author} {\bibfnamefont {J.}~\bibnamefont {Bradbury}}, \bibinfo {author} {\bibfnamefont {R.}~\bibnamefont {Frostig}}, \bibinfo {author} {\bibfnamefont {P.}~\bibnamefont {Hawkins}}, \bibinfo {author} {\bibfnamefont {M.~J.}\ \bibnamefont {Johnson}}, \bibinfo {author} {\bibfnamefont {Y.}~\bibnamefont {Katariya}}, \bibinfo {author} {\bibfnamefont {C.}~\bibnamefont {Leary}}, \bibinfo {author} {\bibfnamefont {D.}~\bibnamefont {Maclaurin}}, \bibinfo {author} {\bibfnamefont {G.}~\bibnamefont {Necula}}, \bibinfo {author} {\bibfnamefont {A.}~\bibnamefont {Paszke}}, \bibinfo {author} {\bibfnamefont {J.}~\bibnamefont {Vander{P}las}}, \bibinfo {author} {\bibfnamefont {S.}~\bibnamefont {Wanderman-{M}ilne}},\ and\ \bibinfo {author} {\bibfnamefont {Q.}~\bibnamefont {Zhang}},\ }\href {https://github.com/jax-ml/jax} {\bibinfo {title} {{JAX}: Composable transformations of python+numpy programs}} (\bibinfo {year} {2018})\BibitemShut {NoStop}%
\bibitem [{\citenamefont {Marquardt}(1963)}]{marquardt1963}%
  \BibitemOpen
  \bibfield  {author} {\bibinfo {author} {\bibfnamefont {D.~W.}\ \bibnamefont {Marquardt}},\ }\bibfield  {title} {\bibinfo {title} {An algorithm for least-squares estimation of nonlinear parameters},\ }\href {https://doi.org/10.1137/0111030} {\bibfield  {journal} {\bibinfo  {journal} {J. Soc. Ind. Appl. Math.}\ }\textbf {\bibinfo {volume} {11}},\ \bibinfo {pages} {431} (\bibinfo {year} {1963})}\BibitemShut {NoStop}%
\bibitem [{\citenamefont {{Materials Project}}(2020)}]{materialsproject_yso}%
  \BibitemOpen
  \bibfield  {author} {\bibinfo {author} {\bibnamefont {{Materials Project}}},\ }\href {https://doi.org/10.17188/1206949} {\bibinfo {title} {Materials data on y2sio5 by materials project}} (\bibinfo {year} {2020}),\ \bibinfo {note} {dataset, material ID mp-3520}\BibitemShut {NoStop}%
\bibitem [{\citenamefont {Ong}\ \emph {et~al.}(2013)\citenamefont {Ong}, \citenamefont {Richards}, \citenamefont {Jain}, \citenamefont {Hautier}, \citenamefont {Kocher}, \citenamefont {Cholia}, \citenamefont {Gunter}, \citenamefont {Chevrier}, \citenamefont {Persson},\ and\ \citenamefont {Ceder}}]{pymatgen2013}%
  \BibitemOpen
  \bibfield  {author} {\bibinfo {author} {\bibfnamefont {S.~P.}\ \bibnamefont {Ong}}, \bibinfo {author} {\bibfnamefont {W.~D.}\ \bibnamefont {Richards}}, \bibinfo {author} {\bibfnamefont {A.}~\bibnamefont {Jain}}, \bibinfo {author} {\bibfnamefont {G.}~\bibnamefont {Hautier}}, \bibinfo {author} {\bibfnamefont {M.}~\bibnamefont {Kocher}}, \bibinfo {author} {\bibfnamefont {S.}~\bibnamefont {Cholia}}, \bibinfo {author} {\bibfnamefont {D.}~\bibnamefont {Gunter}}, \bibinfo {author} {\bibfnamefont {V.~L.}\ \bibnamefont {Chevrier}}, \bibinfo {author} {\bibfnamefont {K.~A.}\ \bibnamefont {Persson}},\ and\ \bibinfo {author} {\bibfnamefont {G.}~\bibnamefont {Ceder}},\ }\bibfield  {title} {\bibinfo {title} {Python materials genomics ({pymatgen}): A robust, open-source python library for materials analysis},\ }\href {https://doi.org/10.1016/j.commatsci.2012.10.028} {\bibfield  {journal} {\bibinfo  {journal} {Comput. Mater. Sci.}\ }\textbf {\bibinfo {volume} {68}},\ \bibinfo {pages} {314} (\bibinfo {year}
  {2013})}\BibitemShut {NoStop}%
\bibitem [{\citenamefont {{National Institute of Standards and Technology}}({\natexlab{b}})}]{nistY}%
  \BibitemOpen
  \bibfield  {author} {\bibinfo {author} {\bibnamefont {{National Institute of Standards and Technology}}},\ }\href {https://physics.nist.gov/PhysRefData/Handbook/Tables/yttriumtable1.htm} {\bibinfo {title} {Atomic data for yttrium ({Y})}},\ \bibinfo {howpublished} {NIST Atomic Spectra Database Handbook} ({\natexlab{b}})\BibitemShut {NoStop}%
\bibitem [{\citenamefont {Abragam}(1961)}]{abragam1961}%
  \BibitemOpen
  \bibfield  {author} {\bibinfo {author} {\bibfnamefont {A.}~\bibnamefont {Abragam}},\ }\href@noop {} {\emph {\bibinfo {title} {The Principles of Nuclear Magnetism}}}\ (\bibinfo  {publisher} {Clarendon Press},\ \bibinfo {address} {Oxford},\ \bibinfo {year} {1961})\BibitemShut {NoStop}%
\bibitem [{\citenamefont {Van~Vleck}(1948)}]{vanvleck1948}%
  \BibitemOpen
  \bibfield  {author} {\bibinfo {author} {\bibfnamefont {J.~H.}\ \bibnamefont {Van~Vleck}},\ }\bibfield  {title} {\bibinfo {title} {The dipolar broadening of magnetic resonance lines in crystals},\ }\href {https://doi.org/10.1103/PhysRev.74.1168} {\bibfield  {journal} {\bibinfo  {journal} {Phys. Rev.}\ }\textbf {\bibinfo {volume} {74}},\ \bibinfo {pages} {1168} (\bibinfo {year} {1948})}\BibitemShut {NoStop}%
\bibitem [{\citenamefont {Isserlis}(1918)}]{isserlis1918}%
  \BibitemOpen
  \bibfield  {author} {\bibinfo {author} {\bibfnamefont {L.}~\bibnamefont {Isserlis}},\ }\bibfield  {title} {\bibinfo {title} {On a formula for the product-moment coefficient of any order of a normal frequency distribution in any number of variables},\ }\href {https://doi.org/10.1093/biomet/12.1-2.134} {\bibfield  {journal} {\bibinfo  {journal} {Biometrika}\ }\textbf {\bibinfo {volume} {12}},\ \bibinfo {pages} {134} (\bibinfo {year} {1918})}\BibitemShut {NoStop}%
\bibitem [{\citenamefont {Anderson}(1954)}]{anderson1954}%
  \BibitemOpen
  \bibfield  {author} {\bibinfo {author} {\bibfnamefont {P.~W.}\ \bibnamefont {Anderson}},\ }\bibfield  {title} {\bibinfo {title} {A mathematical model for the narrowing of spectral lines by exchange or motion},\ }\href {https://doi.org/10.1143/JPSJ.9.316} {\bibfield  {journal} {\bibinfo  {journal} {J. Phys. Soc. Jpn.}\ }\textbf {\bibinfo {volume} {9}},\ \bibinfo {pages} {316} (\bibinfo {year} {1954})}\BibitemShut {NoStop}%
\bibitem [{\citenamefont {Budoyo}\ \emph {et~al.}(2018)\citenamefont {Budoyo}, \citenamefont {Kakuyanagi}, \citenamefont {Toida}, \citenamefont {Matsuzaki}, \citenamefont {Munro}, \citenamefont {Yamaguchi},\ and\ \citenamefont {Saito}}]{budoyo2018}%
  \BibitemOpen
  \bibfield  {author} {\bibinfo {author} {\bibfnamefont {R.~P.}\ \bibnamefont {Budoyo}}, \bibinfo {author} {\bibfnamefont {K.}~\bibnamefont {Kakuyanagi}}, \bibinfo {author} {\bibfnamefont {H.}~\bibnamefont {Toida}}, \bibinfo {author} {\bibfnamefont {Y.}~\bibnamefont {Matsuzaki}}, \bibinfo {author} {\bibfnamefont {W.~J.}\ \bibnamefont {Munro}}, \bibinfo {author} {\bibfnamefont {H.}~\bibnamefont {Yamaguchi}},\ and\ \bibinfo {author} {\bibfnamefont {S.}~\bibnamefont {Saito}},\ }\bibfield  {title} {\bibinfo {title} {Phonon-bottlenecked spin relaxation of {Er$^{3+}$:Y$_2$SiO$_5$} at sub-kelvin temperatures},\ }\href {https://doi.org/10.7567/APEX.11.043002} {\bibfield  {journal} {\bibinfo  {journal} {Appl. Phys. Express}\ }\textbf {\bibinfo {volume} {11}},\ \bibinfo {pages} {043002} (\bibinfo {year} {2018})}\BibitemShut {NoStop}%
\bibitem [{\citenamefont {Welinski}\ \emph {et~al.}(2017)\citenamefont {Welinski}, \citenamefont {Thiel}, \citenamefont {Dajczgewand}, \citenamefont {Ferrier}, \citenamefont {Cone}, \citenamefont {Macfarlane}, \citenamefont {Chaneli{\`e}re}, \citenamefont {Louchet-Chauvet},\ and\ \citenamefont {Goldner}}]{welinski2017}%
  \BibitemOpen
  \bibfield  {author} {\bibinfo {author} {\bibfnamefont {S.}~\bibnamefont {Welinski}}, \bibinfo {author} {\bibfnamefont {C.~W.}\ \bibnamefont {Thiel}}, \bibinfo {author} {\bibfnamefont {J.}~\bibnamefont {Dajczgewand}}, \bibinfo {author} {\bibfnamefont {A.}~\bibnamefont {Ferrier}}, \bibinfo {author} {\bibfnamefont {R.~L.}\ \bibnamefont {Cone}}, \bibinfo {author} {\bibfnamefont {R.~M.}\ \bibnamefont {Macfarlane}}, \bibinfo {author} {\bibfnamefont {T.}~\bibnamefont {Chaneli{\`e}re}}, \bibinfo {author} {\bibfnamefont {A.}~\bibnamefont {Louchet-Chauvet}},\ and\ \bibinfo {author} {\bibfnamefont {P.}~\bibnamefont {Goldner}},\ }\bibfield  {title} {\bibinfo {title} {Effects of disorder on optical and electron spin linewidths in {Er$^{3+}$,Sc$^{3+}$:Y$_2$SiO$_5$}},\ }\href {https://doi.org/10.1016/j.optmat.2016.09.039} {\bibfield  {journal} {\bibinfo  {journal} {Opt. Mater.}\ }\textbf {\bibinfo {volume} {63}},\ \bibinfo {pages} {69} (\bibinfo {year} {2017})}\BibitemShut {NoStop}%
\bibitem [{\citenamefont {Galland}\ \emph {et~al.}(2020)\citenamefont {Galland}, \citenamefont {Lu{\v c}i{\'c}}, \citenamefont {Fang}, \citenamefont {Zhang}, \citenamefont {Le~Targat}, \citenamefont {Ferrier}, \citenamefont {Goldner}, \citenamefont {Seidelin},\ and\ \citenamefont {Le~Coq}}]{galland2020}%
  \BibitemOpen
  \bibfield  {author} {\bibinfo {author} {\bibfnamefont {N.}~\bibnamefont {Galland}}, \bibinfo {author} {\bibfnamefont {N.}~\bibnamefont {Lu{\v c}i{\'c}}}, \bibinfo {author} {\bibfnamefont {B.}~\bibnamefont {Fang}}, \bibinfo {author} {\bibfnamefont {S.}~\bibnamefont {Zhang}}, \bibinfo {author} {\bibfnamefont {R.}~\bibnamefont {Le~Targat}}, \bibinfo {author} {\bibfnamefont {A.}~\bibnamefont {Ferrier}}, \bibinfo {author} {\bibfnamefont {P.}~\bibnamefont {Goldner}}, \bibinfo {author} {\bibfnamefont {S.}~\bibnamefont {Seidelin}},\ and\ \bibinfo {author} {\bibfnamefont {Y.}~\bibnamefont {Le~Coq}},\ }\bibfield  {title} {\bibinfo {title} {Mechanical tunability of an ultranarrow spectral feature of a rare-earth-doped crystal via uniaxial stress},\ }\href {https://doi.org/10.1103/PhysRevApplied.13.044022} {\bibfield  {journal} {\bibinfo  {journal} {Phys. Rev. Applied}\ }\textbf {\bibinfo {volume} {13}},\ \bibinfo {pages} {044022} (\bibinfo {year} {2020})}\BibitemShut {NoStop}%
\end{thebibliography}%

\end{document}